\documentclass[12pt]{iopart}

\usepackage{graphicx}
\usepackage{framed}
\usepackage{color}
\usepackage{nicefrac}

\usepackage{hyperref}  
\hypersetup{unicode,pdftex,colorlinks=true,allcolors=blue}  
\usepackage{hypcap}  
\usepackage{bookmark}   
\bookmarksetup{numbered,open}  

\usepackage{comment}
\includecomment{comment}    
\includecomment{todo} 
\includecomment{comment2} 
\excludecomment{comment2}  

\usepackage{lineno}

\usepackage{braket}
\usepackage[mmddyyyy]{datetime}

\begin{document}

\def\URS{URu$_2$Si$_2$}
\def\TRS{ThRu$_2$Si$_2$}
\def\RS{Ru$_2$Si$_2$}
\def\SiT{Si-terminated }
\def\UT{U-terminated }
\def\ThT{Th-terminated }
\def\etal{et al.}

\def\EF{$E_\mathrm{F}$}
\def\kF{$k_\mathrm{F}$}
\def\vF{$v_\mathrm{F}$}
\def\kpar{$k_\mathrm{||}$}
\def\kB{$k_\mathrm{B}$}
\def\THO{$T_{HO}$}
\def\Tcoh{$T^*$}
\def\invA{\AA$^{-1}$} 
\def\kpar{$k_{\vert\vert}$}
\def\kz{$k_z$}
\def\deg{$^{\circ}$}
\def\kxky{$k_x$-$k_y$ }
\def\kxkz{$k_x$-$k_z$ }
\def\hv{$h\nu$}
\def\Akw{$A(k,\omega)$}
\def\adir{$<$100$>$}
\def\bdir{$<$010$>$}
\def\abdir{$<$110$>$}
\def\cdir{$<$001$>$}
\def\dtof{5$d\rightarrow$5$f$}

\def\G{$\Gamma$}
\def\Gbar{$\overline{\Gamma}$}
\def\Xbar{$\overline{\textrm{X}}$}
\def\Mbar{$\overline{\textrm{M}}$}   
\def\Ybar{$\overline{\textrm{Y}}$}
\def\Sbar{$\overline{\textrm{S}}$}
\def\Mbar{$\overline{\textrm{M}}$}

\def\cred{\color{red}}
\def\cblue{\color{blue}}
\definecolor{dkgreen}{rgb}{0.15,0.35,0.10}
\def\cgreen{\color{dkgreen}}

\title[]{Global perspectives of the bulk electronic structure of URu$_2$Si$_2$ from angle-resolved photoemission}

\author{J. D. Denlinger$^1$, J.-S. Kang$^2$, L. Dudy$^3$, J. W. Allen$^4$,  Kyoo Kim$^5$, J.-H. Shim$^6$,  K. Haule$^7$, J. L. Sarrao$^8$, N.P. Butch$^{9,10}$ and M. B. Maple$^{11}$}

\address{$^1$ Lawrence Berkeley National Laboratory, Berkeley CA 94720 USA}
\address{$^2$ Department of Physics, The Catholic University of Korea, Bucheon 14662, Korea}
\address{$^3$ Synchrotron SOLEIL, L'Orme des Merisiers, 91190 Saint-Aubin, France}
\address{$^4$ Department of Physics, Randall Laboratory, University of Michigan, Ann Arbor, MI 48109, USA}
\address{$^5$ Korea Atomic Energy Research Institute (KAERI), Daejeon 34057, Korea}  
\address{$^6$ Department of Chemistry and Division of Advanced Nuclear Engineering, POSTECH, Pohang 37673, Korea} 
\address{$^7$ Department of Physics and Astronomy, Rutgers University, Piscataway, NJ 08854, USA}   
\address{$^8$ Los Alamos National Laboratory, Los Alamos, New Mexico 87545, USA}
\address{$^9$ NIST Center for Neutron Research, National Institute of Standards and Technology,
100 Bureau Drive, Gaithersburg, Maryland 20899, USA}
\address{$^{10}$Quantum Materials Center, Department of Physics, University of Maryland, College Park, Maryland 20742, USA}  
\address{$^{11}$ Department of Physics, University of California at San Diego, La Jolla, CA 92903, USA}

\ead{jddenlinger@lbl.gov}
\indent
\vspace{10pt}
\date{\today}
\begin{indented}
\item[]\textbf{Keywords:} ARPES, heavy fermions, electronic structure, hidden order
\end{indented}


\begin{abstract}
Previous high-resolution angle-resolved photoemission (ARPES) studies of URu$_2$Si$_2$ have characterized the temperature-dependent behavior of narrow-band states close to the Fermi level (\EF) at low photon energies near the zone center, with an emphasis on electronic reconstruction due to Brillouin zone folding. 
A substantial challenge to a proper description is that these states interact with other hole-band states that are generally absent from bulk-sensitive soft x-ray ARPES measurements. 
Here we provide a more global $k$-space context for the presence of such states and their relation to the bulk Fermi surface topology using synchrotron-based wide-angle and photon energy-dependent ARPES mapping of the electronic structure using photon energies intermediate between the low-energy regime and the high-energy soft x-ray regime.  
Small-spot spatial dependence, $f$-resonant photoemission, Si 2$p$ core-levels, x-ray polarization, surface-dosing modification, and theoretical surface slab calculations are employed to assist identification of bulk versus surface state character of the \EF-crossing bands and their relation to specific U- or Si-terminations of the cleaved surface.  
The bulk Fermi surface topology is critically compared to density functional theory and 
to dynamical mean field theory calculations. 
In addition to clarifying some aspects of the previously measured high symmetry \G, Z and X points,
incommensurate 0.6a* nested Fermi-edge states located along Z-N-Z are found to be distinctly different from the density functional theory Fermi surface prediction.
The temperature evolution of these states above \THO, combined with a more detailed theoretical investigation of this region, suggests a key role of the N-point in the hidden order transition.
\end{abstract}

01/20/2022
%
%
%
%
%

\
\tableofcontents

\newpage

\URS\ is a paradigm heavy fermion compound exhibiting a distinct temperature dependent experimental phase transition below 17.5 K with an accompanying loss of entropy as exhibited by specific heat \cite{Palstra1985,Maple1986,Schlabitz1986}, and other key transport properties that point to a significant gapping of the Fermi surface (FS).   
The nature of this unconventional ordered state has been a 35-year focus of research generating many proposed theoretical models, and extensive experimental elucidation of phase diagrams as a function of temperature, pressure, magnetic field and chemical substitutions in order to isolate and understand the existence of the unknown ``hidden order'' (HO) phase relative to other neighboring low temperature large moment anti-ferromagnetic (LMAF), ferromagnetic, and superconducting phases (see review articles \cite{Mydosh2011,Mydosh2014,Mydosh2020}).
Key HO properties relevant to this electronic structure study include the observation of commensurate and incommensurate spin-excitation momentum vectors by inelastic neutron scattering \cite{Wiebe2007},
and a zone-folded lowering of the electronic symmetry of the high temperature paramagnetic (PM) phase electronic states in both the HO and LMAF phases  \cite{Hassinger2010}. 

In principle, angle-resolved photoemission (ARPES) can 
(i) identify the $k$-resolved bulk FS topology, 
(ii)  identify $k$-locations of FS nesting vectors, and 
(iii) observe temperature ($T$) dependent changes including the evolution of Kondo hybridization and HO gapping and zone-folding.
Previous low photon energy high-resolution ARPES studies have focused on the $T$-dependent appearance of a very narrow ``M''-shaped band ($M$-band) feature close to \EF\ at the zone center in the HO phase which is absent in the paramagnetic (PM) phase, and is postulated to arise from zone-folding associated with a PM body-centered tetragonal (bct) to HO tetragonal Brillouin zone change \cite{Santander2009,Yoshida2010,Chatterjee2013,Boariu2013,Yoshida2013,Bareille2014}.
Also at higher photon energies, lower resolution but bulk sensitive SX-ARPES has identified a PM phase bulk FS topology \cite{Kawasaki2011,Fujimori2021}, which exhibits both key discrepancies with the density functional theory (DFT) FS predictions and does not observe features of the low energy ARPES either due to lack of resolution or to unclarified surface state  (SS) origins.

In this work we address these outstanding electronic structure issues using ARPES in intermediate photon energy ranges with wide multi-BZ momentum-space coverage, sufficient energy resolution to perform $T$-dependence studies, and with detailed cleave surface characterization to identify surface states and elucidate the bulk-like features in this surface sensitive regime.  
Along the way, in bridging this photon energy gap in the \URS\ ARPES literature, a topical review of the previous ARPES studies and interpretations is provided.

\section{Introduction} \label{intro}

\subsection{ARPES $k$-space guide}

\sectionmark{FIG. 1. INTRO }
\begin{figure}[t]
\begin{framed}
\begin{center}
\includegraphics[width=15cm]{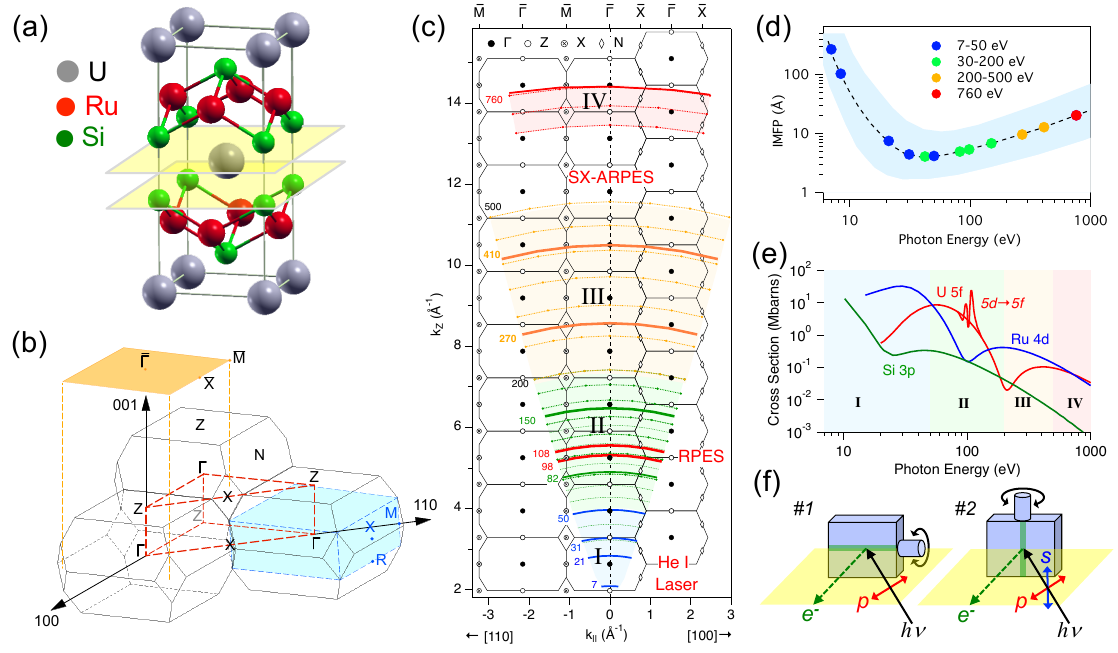}
\caption{
\textbf{Momentum space guide for ARPES.}   
(a) Body-centered tetragonal (bct) crystal structure of \URS\ highlighting the atomic layering of atoms and the known cleavage planes between U and Si layers. 
(b) Three-dimensional view of the momentum space bct Brillouin zone illustrating the staggered stacking of BZs in the \adir-\cdir\ plane, multi-BZ high symmetry cuts and the projected surface BZ. 
Also the tetragonal zone-folded BZ of the ordered phase is illustrated (dashed, blue-shaded).
(c) Multi-BZ momentum space guide for illustrating the various photon energy ranges of this study and other literature studies spanning five different photon energy ranges (Roman numerals). 
Specific fixed photon energy arcs of interest are identified.  An inner potential of 14 eV has been used for the schematic.
(d) Inelastic mean free path, i.e. exponential attenuation length, of Fermi-edge photoelectrons with kinetic energy equal to the photon energy minus the surface work function ($\sim$4.5 eV).  The range of material dependent variations is indicated in blue.
(e) Photoionization cross section of relevant near \EF\ U 5$f$, Ru 4$d$, and Si 3$p$ orbitals.
 (f) Two experimental geometries used in the ARPES measurements with different electron analyzer slit orientations and orthogonal sample scan rotation axis, and different incident x-ray (s,p) polarization capabilities.
}
\label{bz}
\end{center}
\end{framed}
\end{figure}

Figure \ref{bz}(a) shows the body-centered tetragonal (bct) crystal structure of \URS\ that consists of atomically layered planes in the sequence of -U-Si-Ru-Si- along the c-axis.  
Also illustrated are the natural cleavage planes between weaker-bonded U and Si layers, a property that is similarly documented in other rare-earth and actinide ruthenium, rhodium and iridium silicides, including most notably the literature on YbRh$_2$Si$_2$ \cite{Guttler2014}. 
While a $single$ natural c-axis cleavage plane is a simplifying factor  
and benefit for generating flat surfaces for ARPES, it still results in the complexity of two possible complementary U- or Si- cleave terminations which is an important topic in this paper. 

ARPES probes the momentum space electronic structure, so it is important to understand the bct Brillouin zone (BZ) and the impact of its multi-BZ stacking, illustrated in figure \ref{bz}(b),  on the high symmetry cuts in $k$-space.
Along normal emission, \G\ and Z points are probed with varying photon energy excitation with -\G-Z-\G- periodicity of 4$\pi$/c = 1.31 \invA. 
In the \adir-\cdir\ plane, the staggered stacking of BZs provides an alternate \kpar\ path between \G\ and Z points in neighboring zones with 2$\pi$/a = 1.52 \invA\ distance.
In the diagonal (110)-(001) plane, high symmetry \kpar\ cuts have simpler -\G-X-\G\- or -Z-X-Z- periodicity of $\sqrt{2}\pi$/a =  2.15 \invA.
Also noteworthy from figure \ref{bz}(b) is that the X-points repeat twice as fast as \G\ or Z points along (001), albeit with a 90$^{\circ}$ rotation between each X-point. 

Additionally shown in figure \ref{bz}(b) are the bulk simple tetragonal BZ (blue shaded, dashed line border), and the projected surface BZ.  
The smaller tetragonal BZ is relevant for the case of the HO and LMAF phases in which alternating U planes become inequivalent.
In the resulting electronic zone-folding, the bct BZ \G\ and Z points become equivalent and the two-fold symmetry of the X-points become 4-fold symmetric.  
The surface BZ projection, relevant to surface states, surface sensitive \kz-broadening and theoretical slab calculations, exhibits similar zone-folding effects that arise from the equivalence of states along \kz. 
Note that the surface BZ \Mbar\ and \Xbar\ and tetragonal BZ M and X point labels (confusingly) correspond to the X-point  and (\G-Z)/2 mid-point of the bct BZ, respectively. 

For an ARPES $k$-space navigation guide, figure \ref{bz}(c) provides a schematic of many bct BZs stacked along the (001) direction illustrating different photon energy ranges measured in the literature and in this study. 
Both \adir-\cdir\ and \abdir-\cdir\ plane BZ stackings are represented, and a 14 eV inner potential barrier at the surface is used to convert the measured Fermi-edge kinetic energies into the \kz\ values \textit{inside} the crystal. 
The conical edges for the shaded regions correspond to a $\pm$15$^{\circ}$ angular detection window typical of imaging electron analyzers, illustrating how different parts of reciprocal space are accessed in different ARPES experiments.

Two additional important factors when comparing ARPES of \URS\ over such a broad energy range are the probe depth of the photoelectrons and the relative photoionization cross section of the U 5$f$, Ru 4$d$ and Si 3$p$ states contributing to the Fermi-edge region \cite{Fujimori2016}.
The surface sensitivity of the ARPES technique is illustrated in the figure \ref{bz}(d) plot of the 
inelastic mean free path (IMFP) \cite{Seah1979,Tanuma2011} of Fermi-edge photoelectrons, which reflects the exponential attenuation length of spectral weight intensity as the photoelectrons travel to the surface.   While probe depths of $>$15 \AA\ can be achieved at the highest and lowest energies, the IMFPs for the ARPES measurements here are less than one $\sim$10 \AA\ c-axis unit cell.
The complex variation of U 5$f$ and Ru 4$d$ cross sections \cite{YehLindau1985} shown in figure \ref{bz}(e)  arising from different Cooper minima (dipole channel interference), results in a dominance of U 5$f$ weight at \EF\ in Regions II and IV, including the extreme 5$f$ enhancement at the U \dtof\ resonances at 98 and 108 eV.  In Region I, the U 5$f$ cross section becomes progressively weaker to lower photon energies, and U 5$f$ character is only inferred from the narrowness of the bands.

Early ARPES studies included laboratory-based He I (21.2 eV) excitation \cite{Ito1999} and a comparative synchrotron ARPES study of  La/Ce\RS\ and Th/U\RS\ \cite{Denlinger2001}  providing moderate resolution mapping of the paramagnetic phase of the \URS\ band structure from $h\nu$=14-228 eV along normal emission, and select high symmetry angle-mapping including the 98 and 108 eV U \dtof\ resonance energies. 
Beginning in 2009, high-resolution $T$-dependent ARPES measurements of \URS\ employing more advanced imaging Scienta analyzers were performed using He I excitation by Santander-Syro \etal\ \cite{Santander2009} and using 7 eV laser excitation by Yoshida \etal\
\cite{Yoshida2010}. This was then followed up by high-resolution synchrotron studies by multiple research groups \cite{Chatterjee2013,Boariu2013, Yoshida2013, Bareille2014} using the same BESSY ``$1^3$''  ARPES endstation with sample cooling to 1K, primarily using photon energies of 19, 31-34, and 50 eV close to zone center high symmetry points.
Complementary lower resolution, but bulk-sensitive, soft x-ray ARPES at high photon energy (680-760 eV) was performed by Kawasaki \etal\ \cite{Kawasaki2011}.  
In addition, time-resolved pump-probe experiments at $\sim$30 eV  \cite{Dakovski2011} and other low energy (14-34 eV)  and $f$-resonant  98 eV ARPES were performed \cite{Meng2013}.  
Reviews of these ARPES studies of \URS\ up to 2014 have been provided by Durakiewicz \cite{Durakiewicz2014}, by Fujimori within a 4$f$/5$f$ ARPES topical review \cite{Fujimori2016}, and also within the overall \URS\ reviews by Mydosh \etal\ \cite{Mydosh2011,Mydosh2014,Mydosh2020}.

While ARPES in the lowest and highest energy Regions labeled I and IV in figure \ref{bz}(c) is well represented by this literature, there is a lack of representation of ARPES mapping at the intermediate photon energies from 50-500 eV. 
Hence in this study we provide a more global $k$-space perspective of ARPES of \URS\ with overview photon- and angle-dependent Fermi surface and band mapping in Regions II and III. 
First in Section \ref{term} we revisit the ARPES of Region II using a high resolution imaging electron spectrometer, and make use of the 98 eV$f$-resonance condition as well as the Si 2$p$ core-level to characterize the cleaved surface termination domain structure and discuss the identification of surface states with assistance from theoretical surface slab calculations \cite{Denlinger2009}.
In Section \ref{smod} surface modifications, including aging, alkali adsorption and atomic-H dosing, supplement this surface state characterization \cite{Denlinger2010}.
We then provide in Section \ref{arpes200},  global mapping of \URS\ in the $<$200 eV Region II  with comparison to the $f^0$ reference compound \TRS, including a focus on a 150 eV high symmetry cut  \cite{Denlinger2009}. Here a new dynamical mean field theory (DFT+DMFT) band calculation with renormalized$f$-bands is introduced for better understanding of the heavy effective mass regions that ARPES is aiming to measure.  
In Section \ref{sxarpes}, we briefly explore ARPES with higher photon energies $>$200 eV in the semi-bulk sensitive Region IV for comparison to the $>$680 eV literature SX-ARPES Region IV results with discussion of resolution-limited patterns of heavy mass regions of the Fermi surface.

We then return to the lower photon energy and higher energy resolution Region II below 98 eV in Section \ref{tdep}.
First we focus on the normal emission \Gbar\ region in Section \ref{gzpt} with photon-dependent variable x-ray polarization mapping to provide a more global context of the \kz\ locations of previous high-resolution measurements of the $M$-band feature  at \G\ and Z points.
Then at the X-point in Section \ref{xpt},  the separation of  a strong  5$f$ surface state
and a shallow bulk f electron pocket is presented with assistance from variable x-ray polarization and $T$ dependence.
In Section \ref{npt}, a new region of interest near the (001) zone-boundary N-point is revealed from global photon-dependent FS mapping, and a theoretical reexamination identifies an additional heavy band mass region just below \EF\ along Z-N-Z. 
Distinctive temperature dependence of the heavy band mass states in this region
 provides evidence for a Kondo hybridization evolution that is interrupted just  \textit{above} \THO\ \cite{Denlinger2014}.
Finally we provide discussion of ordered phase zone-folding, heavy mass origins, the itinerant versus localized debate, nesting vectors, FS gapping and pseudogaps, as they relate to this newly identified N-point region that has rich potential for explanation of  the hidden order phenomena.

\section{Methods}

\indent\indent
\textbf{Crystals.}
Single crystals of \URS\  were synthesized from a polycrystalline boule utilizing a tri-arc furnace equipped with a Czochralski crystal puller \cite{Jeffries2008}. 
\TRS\ crystals were obtained from the formation of small crystalline platelets from the surface of a rapidly quenched polycrystalline melt on a copper hearth.

\textbf{Sample surface.}
Crystals were cleaved in-vacuum at $T\sim$20 K using the top-post method exposing the (001) surface. 
For surface modifications in Section \ref{smod}, adsorption of potassium atoms was achieved by heating of a commercial SAES alkali metal dispenser in front of the sample surface, using fixed 1 min exposure times (with variable flux) between measurements.  Atomic-H surface dosing was achieved by the thermal cracking of H$_2$ gas passed through  an e-beam heated tungsten capillary. 
 
\textbf{ARPES.}
 Experimental measurements in the 80-200 eV and 200-500 eV regions, presented in Sections \ref{term}, \ref{arpes200} and  \ref{sxarpes}  employed the 2009 configuration of Beamline 7.0 of the Advanced Light Source (ALS), including a Scienta R4000 imaging electron analyzer with 0.2$^{\circ}$ angular resolution, and linear horizontal (LH, $p$-) polarization only of the incident x-rays.  
Dual-angle mapping with a horizontal analyzer slit was performed by fine scanning of an orthogonal tilt stage (geometry \#1 in figure \ref{bz}(f)), and mapping of different regions of the sample surface was achieved with sample XY compensation correlated to the tilt motion.
Temperature-dependent measurements in the 30-150 eV region presented in Section \ref{tdep} employed the 2012 ARPES configuration of the ALS Beamline 4.0.3 which has additional polarization control from the undulator source including linear vertical (LV, $s$-) polarization.  
Dual-angle mapping with a vertical analyzer slit was achieved by polar rotation of the entire XYZ manipulator (geometry \#2) which allows positioning of any point of the sample surface to the polar rotation axis. 
In this manuscript, valence band dispersion $E(k)$ spectral intensity images 
are referred to as ``VB maps'', and Fermi-edge spectral intensity image slices from dual-angle or photon-dependent data sets are referred to as ``FS maps".

\textbf{Theory calculations.}
Density functional theory electronic structure calculations including surface slab calculations of \TRS\ presented in Section \ref{spatial} and bulk calculations of \TRS\ and \URS\  in Section \ref{arpes200} were performed using WIEN2k \cite{Blaha2020} using the generalized gradient approximation (GGA). 
Minimal two-cell surface slab structures of \TRS\ were used without surface relaxation from bulk sites and without spin-orbit coupling (SOC). 
The all-electron DFT code allowed the simulation of the $\sim$100 eV binding energy {\cblue BE} Si 2$p$  core-level with suitable choice of the core energy cutoff.
To address U 5$f$ state interaction with Si-terminated surface states, a similar two-unit cell slab calculation was performed employing DFT+DMFT \cite{Kotliar2006}, 
and also two- and three-unit cell DFT slab calculations of the \URS\ X-point are presented in section \ref{xss}. 
Additionally, a bulk DFT+DMFT calculation of high symmetry bands is introduced in Section \ref{urs200} for comparison to experimental band dispersions.
The charge-self-consistent bulk DFT+DMFT calculation using onsite Coulomb correlation energy  $U$=4 eV, and the $f$-itineracy was tuned via the double-counting parameter in the 
OCA impurity solver  resulting in an $f$-occupation of  $n_f$=2.4 \cite{Kung2015}, and a 5$\times$ $f$-bandwidth renormalization relative to DFT. 
The crystalline electric field (CEF) states were not treated beyond the DFT input.

\section{Surface termination} \label{term}
\subsection{Spatial characterization} \label{spatial}

Despite the known natural cleavage plane and scanning tunneling microscopy (STM) characterization of cleaved \URS\ surfaces \cite{Aynajian2010}, there had been no published discussion of different surface terminations in \URS\ ARPES until a recent combined ARPES and STM study by Zhang \etal\ \cite{Zhang2018}.  The He I ARPES measurements were able to distinguish predominantly U and Si terminations from \textit{different} cleaved surfaces using a 1 mm beam spot, most notably observing different hole-like versus electron-like dispersions at the X-point. 
In the studies presented here, $every$ ARPES experiment, after coarse spatial and angular alignment, begins with a fine spatial characterization 
of the sample surface domain structure with 50 $\mu$m x-ray spot resolution via rastering of the sample XY position. 
Two methods for rapid and high contrast domain characterization include exploiting (i) the large dynamic range contrast of U \dtof\ on- and off-resonance energies, and also (ii) energy shifted components of the Si 2$p$ core-level lineshape. 

\sectionmark{FIG. 2. XY }
\begin{figure}[t]
\begin{framed}
\begin{center}
\includegraphics[width=15cm]{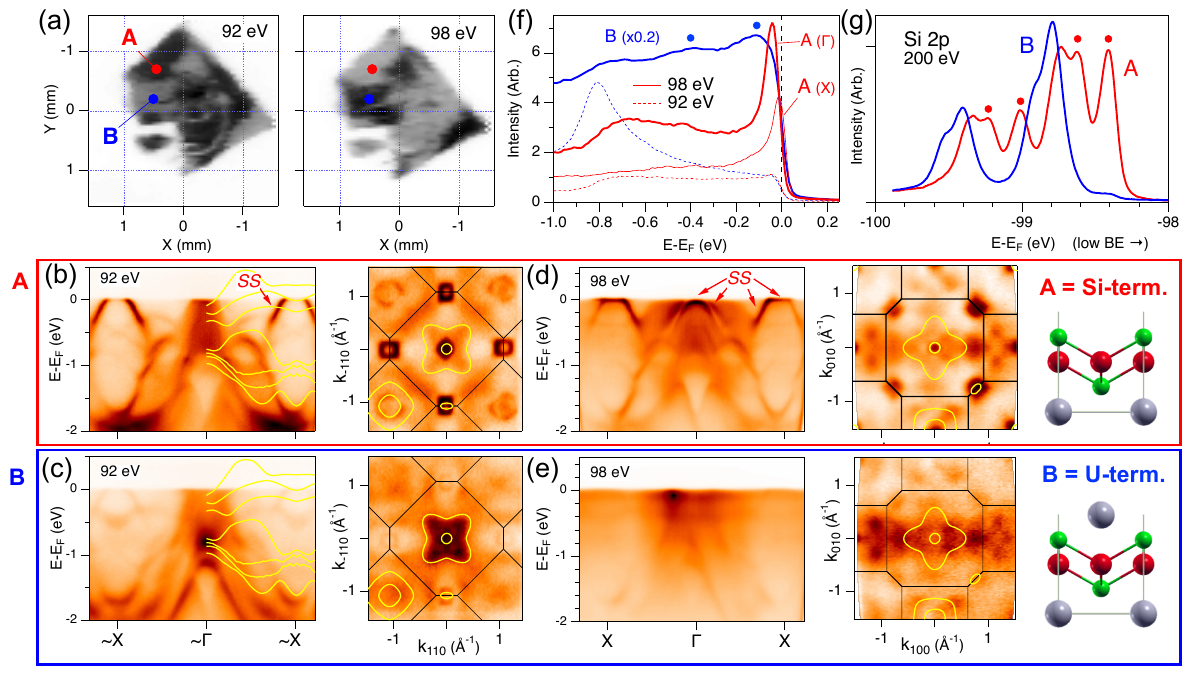}
\caption{
\textbf{Cleave surface spatial characterization.} 
(a) Fermi-energy intensity XY maps of a typical cleaved surface of \URS\ for off-resonance 92 eV and on-resonance 98 eV excitation energies where non-$f$ and 5$f$ spectral weight are respectively highlighted, and reveal two complementary surface regions, and representative points labeled  A and B. 
(b,c) Off-resonance normal emission 92 eV VB and FS maps for points A and B, with bulk DFT bands and FS contours overplotted (yellow).
(d,e) On-resonance normal emission 98 eV VB and FS maps for points A and B.
(f) Comparison of off- and on-resonance valence band spectra for points A and B showing differences in the $f$-weight lineshape.
(g) Comparison of the Si 2$p$ core-level lineshape for points A and B measured with 200 eV excitation.
}
\label{xy}
\end{center}
\end{framed}
\end{figure}

Figure \ref{xy}(a) shows Fermi-edge intensity XY maps of a typical cleave surface at both the on-resonance 98 eV excitation and also an off-resonance 92 eV photon energy.  
Strikingly, the two spatial images show sharp domain boundaries between high and low intensity and with a very complementary reversal of domain intensity between the two excitation energies. 
Logically, the high intensity of the 98 eV map is expected to correspond to the U-terminated surface, and the lower intensity region  is expected to correspond to  the Si-termination where the topmost U atoms are buried below a surface -Si-Ru-Si triple layer.
The inverse complementary high contrast domain structure of the 92 eV spatial map where non-$f$ spectral weight dominates is explained by the comparison of VB maps at points A and B in the two different domain regions, as shown in figure \ref{xy}(b) and (c) along X-\G-X. 
For off-resonance excitation at 92 eV, the standout feature is the strong hole pockets at X for point A, in contrast to a very weak presence at point B. 
Additionally the hole like intensities at \G\ for point A are replaced by a high density of states (DOS) point at -0.8 eV with an emerging electron-like band dispersion to \EF.
Off-resonance 92 eV FS maps at points A and B in figure \ref{xy}(d,e) additionally highlight this extreme X-point surface termination difference, but with residual weaker intensities of B in the FS map of point A (and vice versa).

For the on-resonance energy of 98 eV, while the overall intensity is weaker for point A, the clarity of the band structure is much greater and includes a strong $f$-character in a narrow peak \textit{at} \EF\ interior to the X-point hole pockets, and at a hole band maximum 30-40 meV \textit{below} \EF\ at \G.
These $k$-dependent $f$-peaks are absent at point B, and are replaced by a strong $k$-independent $f$-spectral weight component and  tail that extends below -1 eV which tends to obscure the higher binding energy (BE) bands.   
The angle-averaged 98 eV spectrum  in figure \ref{xy}(f) from point B reveals finite BE peaks at -0.1 and -0.4 eV, perhaps reflective of a localized character of $f$-states of the surface U atoms.

It is also apparent from the valence band images that optimization of the sample position at off-resonance photon energies based on the intensity, sharpness and clarity of dispersing bands (observed live for the imaging spectrometers) will naturally select the Si terminated surface.  
Indeed common to all published high-resolution low photon energy studies of \URS\ is the presence of the intense $\sim$30 meV hole band maximum at normal emission, a signature of the \SiT surface, which was initially identified as a surface state in the early high-resolution He I study by Santandar-Syro \etal\ \cite{Santander2009}. 
Similar to the 98 eV point B results here, the recent He I study by  Zhang \etal\ observed a greatly diminished intensity of this 30 meV surface state for the \UT surface \cite{Zhang2018}.

An alternate rapid spatial characterization of surface domains, available for photon energies greater than 100 eV, comes from the comparison of the Si 2$p$ core-level lineshape in figure \ref{xy}(g), which includes a 0.6 eV SO-splitting of $j$=($\frac{3}{2}$,$\frac{1}{2}$) components.
The standout feature is the strong low BE shifted peaks for point A that are only weakly present for point B.   
Such a low BE component is generally consistent with photoemission spectroscopy of clean Si single crystal surfaces where the origin is that of surface adatoms with an unfilled dangling bond state \cite{Himpsel1980}. 
The weak $<$5\% amplitude of the low BE Si 2$p$ peak in the point B data, indicating a  small percentage presence of Si-termination regions within the 50 $\mu$m beam spot, is consistent with the residual appearance of the X-point square contours and hole bands in the 92 eV FS map and spectrum of point B.
Surface termination domain mixtures on the sub-micron scale are evident in STM line-scan step heights between terraces  
\cite{Aynajian2010,Zhang2018}. 

Comparison to theoretical bulk bands is also a standard first method for identifying surface states.
The strong Si-terminated states near \EF\ have no counter part in the bulk DFT bands or FS contours, and are therefore inferred to be surface states.  
A convergence of theory \G-point bands to a -0.8 eV high DOS peak for point B at 92 eV, and agreement of \G-plane DFT FS contours with some shapes in the 92 eV map supports the assumption that the U-terminated surface is more bulk-like, albeit with a potential disadvantage of less clarity of bands from the surface U $f$-weight, at least for the on-resonance condition.  Also the apparent DFT agreements to the 92 eV FS map are not observed in the on-resonance map of Fermi-edge $f$-weight.

\subsection{Theoretical surface slab calculations}  \label{dftslab}

\sectionmark{FIG. 3. DFT SLAB }
\begin{figure}[t]
\begin{framed}
\begin{center}
\includegraphics[width=15cm]{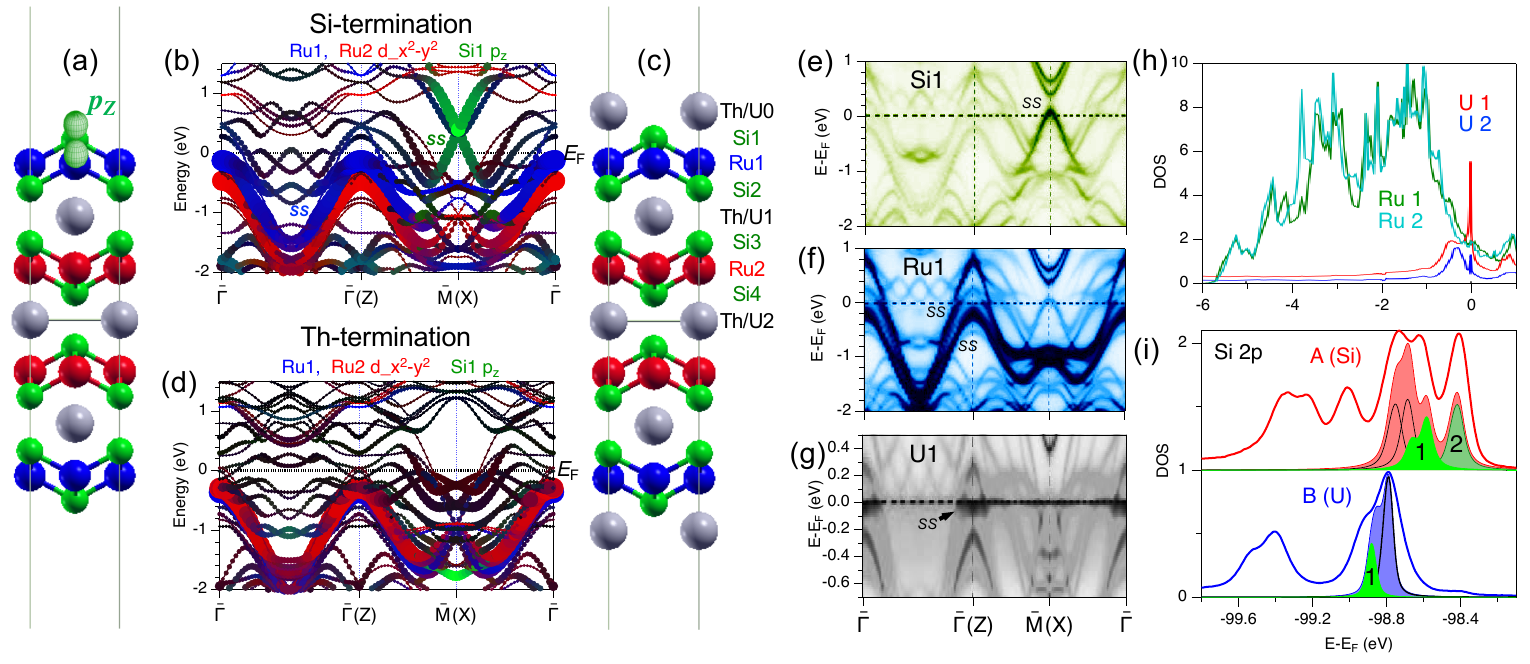}
\caption{
\textbf{Theoretical surface slab calculations of \TRS\ and \URS\ for two different surface terminations.} 
(a,b) Si-terminated 15-layer slab structure of \TRS\ and corresponding band structure.
(c,d) Th-terminated 17-layer slab structure of \TRS\ and corresponding band structure. 
Color-coded band highlighting is indicated: (red) bulk-like Ru2-d, (blue) near surface Ru1-d, and (green) topmost Si1-$p_z$ orbital character.
The $\sim$1 bulk unit cell of vacuum space in the periodic slab unit cell is indicated by the length of the vertical lines.
The surface dangling bond Si $p_z$ orbital is highlighted in green in (d). 
(e,f,g) DFT+DMFT spectral functions calculated for a Si-terminated \URS\ slab for (e) top Si1 site, (f) second layer Ru1 site, and (g) fourth layer U1 site.
(h) Partial DOS for the two Ru and U sites showing an enhanced $f$-peak at \EF\ for the U1 site.
(i) Si 2$p$ core-level spectra with comparison to slab calculations of the relative energy shifts for the $j$=$\frac{3}{2}$ components. Numbered green components correspond to the top two Si sites (Si1, Si2).
}
\label{slab}
\end{center}
\end{framed}
\end{figure}

While the identification of the Si-termination surface states can be confidently made from comparison to \UT spectra and to bulk DFT theory (and also by their response to surface modification described in Section \ref{smod}), it is useful to theoretically reproduce these experimental surface termination differences using surface slab calculations and gain deeper insight into the \textit{origins} of the surface states from their atomic site and orbital character. 
A recent DFT+DMFT surface slab calculation presented by Zhang \etal \cite{Zhang2018} confirmed the X-point hole-pocket versus electron-like band dispersion reversal between the two different surface terminations.  However, it failed to reproduce the dominant intensity 30 meV surface state hole band for the \SiT surface, because only one unit cell 7- and 9-layer slab structures were used, which contain only one near-surface Ru site and no bulk-like Ru site for reference. 

Figure \ref{slab}(a) and (c) show two $\sim$2 unit cell Si- and \ThT periodic surface slab structures of \TRS\ consisting of 15- and 17-layers of atoms and $\sim$1 unit cell of vacuum. 
The electronic band structures of the Si- and \ThT slabs  are shown in figure \ref{slab}(b) and (d), respectively, with color-coded enhancement of the topmost Si site $p_z$ character and of the bulk and near surface Ru site $d_{x^2-y^2}$-character. 
The first striking result for the \SiT band structure is the emergence of Dirac-like linear dispersing bands with a crossing point +0.5 above \EF\  in the middle of an X-point bulk-projected band gap region for the \ThT band structure.
The distinct Si 2$p_z$ orbital character of the Dirac-like band confirms its identity as the Si dangling bond surface state.
The lower part of the band structure nicely agrees with the X-point hole-pocket \EF-crossing observed in ARPES.
In contrast, the \ThT case is observed to have the opposite electron-like dispersion to \EF\ at X with Ru $d$-character including a similar linear dispersion but with a crossing point at -1 eV instead of +0.5 eV.
The one-unit cell slab calculation of Zhang et al. also produces an artificial
energy splitting of the X-point hole band into two separate ($p_z$\ and $p_x$+$p_y$) bands, due to the top and bottom Si surface state wave functions being allowed to interact with each other through the too-thin 7-layer \SiT slab structure.

Also notable is that while the \ThT case has a strong Ru 5$d_{x^2-y^2}$-band dispersion between -0.5 and -2 eV, the removal of the surface Th atoms shifts this band to higher energy with a maximum very close to \EF\ at \G, $and$ generates similar dispersing split-off bands that form a hole-crossing at \G.  
This \textit{combination} of hole pocket and hole band maximum just below \EF\ at \G\ has a strong resemblance to the on-resonance 98 eV spectrum for Si-termination shown in figure \ref{xy}(d), as well as to all the high-resolution low photon energy ARPES studies. 
The surface state origin of the outer hole band has not been previously discussed, 
except for its notable absence in high energy bulk-sensitive SX-ARPES \cite{Kawasaki2011}. 
Similar surface-termination differences are also observed in the measurement and analysis of YbCo$_2$Si$_2$ \cite{Guttler2014}, including a Si-termination surface state in the X-point (\Mbar) bulk projected gap region (with a crossing point \textit{below} \EF), and a Si-termination \Gbar\ hole band shifted closer to \EF\ than its bulk counterpart.

Since the \TRS\ theory slab calculations above inherently do not explore any interaction of the U 5$f$ states with the Si-terminated Si $p_z$ or Ru $d$-character surface states, 
  figure \ref{slab}(e),(f) and (g) show the \Akw\ spectral function band structures for the topmost Si1, Ru1 and U1 sites,  obtained from a DFT+DMFT calculation of the same two unit cell slab structure.
The top layer Si1 \Akw\ confirms the surface origin of the X-point (\Mbar) linear band hole pocket (with spin-orbit coupling causing gapping of the crossing point above \EF), but also with some spectral weight participation from the Ru1, Si2 (not shown) and U1 orbitals. 
At \Gbar, the second layer Ru1 \Akw\ confirms the near surface origin of the outer hole band that crosses \EF\ and extends to +0.8 eV, and which contains negligible character from the interior Ru2 site (not shown).
Most interestingly, the fourth layer U1 site \Akw\ shows a narrow band of enhanced $f$-weight just below \EF\ at \Gbar\  which is also absent  of any spectral weight contribution from the central bulk-like U2 site (not shown). 
While the confinement of this narrow band interior to the outer hole-band surface state is suggestive of the experimental result at the X-point,
its $M$-shaped dispersion is reminiscent of the heavily studied normal emission narrow band in multiple low-energy high-resolution ARPES studies \cite{Santander2009,Yoshida2010,Chatterjee2013,Boariu2013, Yoshida2013, Bareille2014}.
The near-surface origin of the narrow band is also reflected in the integrated density of states (DOS) plot in figure \ref{slab}{h}, where the U1 DOS exhibits a very strong narrow peak at \EF\ that is much weaker for the interior U2 site DOS.
However, this theory slab result of a narrow $M$-band at the surface of the PM phase bulk, does not prove that the experimental HO phase $M$-band is actually a surface state.   Rather, it merely reflects the similarity of zone-folding effects of the bct BZ in such slab calculation projections to the surface BZ as compared to the bulk tetragonal BZ in the HO state.
The $M$-band is further discussed in section \ref{zonefold} and the X-point U 5$f$ surface state(s)
are investigated in detail in Section \ref{xpt} including additional surface slab calculations. 

\textbf{Si 2$p$ core level.} Also from the \TRS\ slab calculations, theoretical simulations of the Si 2$p$ core-level lineshapes for the two surface terminations, shown in figure \ref{slab}(i) agree well with the photoemission spectra.
However, for the \SiT\ surface, an unexpected  revelation is that the -0.25 eV lowest binding energy shifted surface state peak actually originates from the \textit{third} layer Si2 site, while the topmost Si1 site with the dangling bond is shifted only -0.1 eV 
relative to the bulk-like energies of the Si sites near the slab center. 
Also evident in the Si1 lineshape is a prediction of a sub-orbital splitting between the in-plane $p_{x,y}$ states and the out-of-pane $p_z$ dangling bond state.
For the \ThT\ surface, the topmost Si1 energy is shifted oppositely to \textit{higher} BE relative to the bulk states consistent with a photoemission high BE shoulder.

\bigskip
At this point of accrued experimental and theoretical understanding of the \SiT surface states, we can discuss the various possible approaches for achieving the experimental goal of isolating the bulk electronic structure of \URS.
With the assumption that the \UT surface represents the truest bulk-like electronic structure, approaches for the suppression of the undesired Si surface states include: 
(i) cleaving many samples to find the highest quality \UT regions with weakest Si surface intensities,
(ii) artificial surface modification to selectively suppress the Si surface states, and
(iii) measurement in a more bulk-sensitive  higher photon energy regime.
Surface modification is explored in the next Section \ref{smod} and  soft x-ray ARPES measurements are discussed in Section \ref{sxarpes}. 

It is noteworthy that the strong broad $k$-independent $f$-weight of the \UT surface is also an undesired non-bulk surface state arising from the less-coordinated surface U atoms, that can contribute a high background that compromises the clarity of features to be measured.   
Thus other experimental approaches to suppress this unwanted surface $f$-weight include: 
 (i) measurement at low photon energy where the U 5$f$ photoionization cross section is low,
 (ii) use of a smaller $<$10 micron beam spot to better isolate higher quality \UT regions with less surface U disorder, and  
(iii) specifically select \SiT regions but measure bulk-related features away from known Si surface states. 
The last of these is a choice made for the optimal sharpness of bulk-like bands in $T$-dependent measurements presented in Section \ref{npt}.

\subsection{Surface modification} \label{smod}

\sectionmark{FIG. 4. DOSE }
\begin{figure}[t]
\begin{framed}
\begin{center}
\includegraphics[width=15cm]{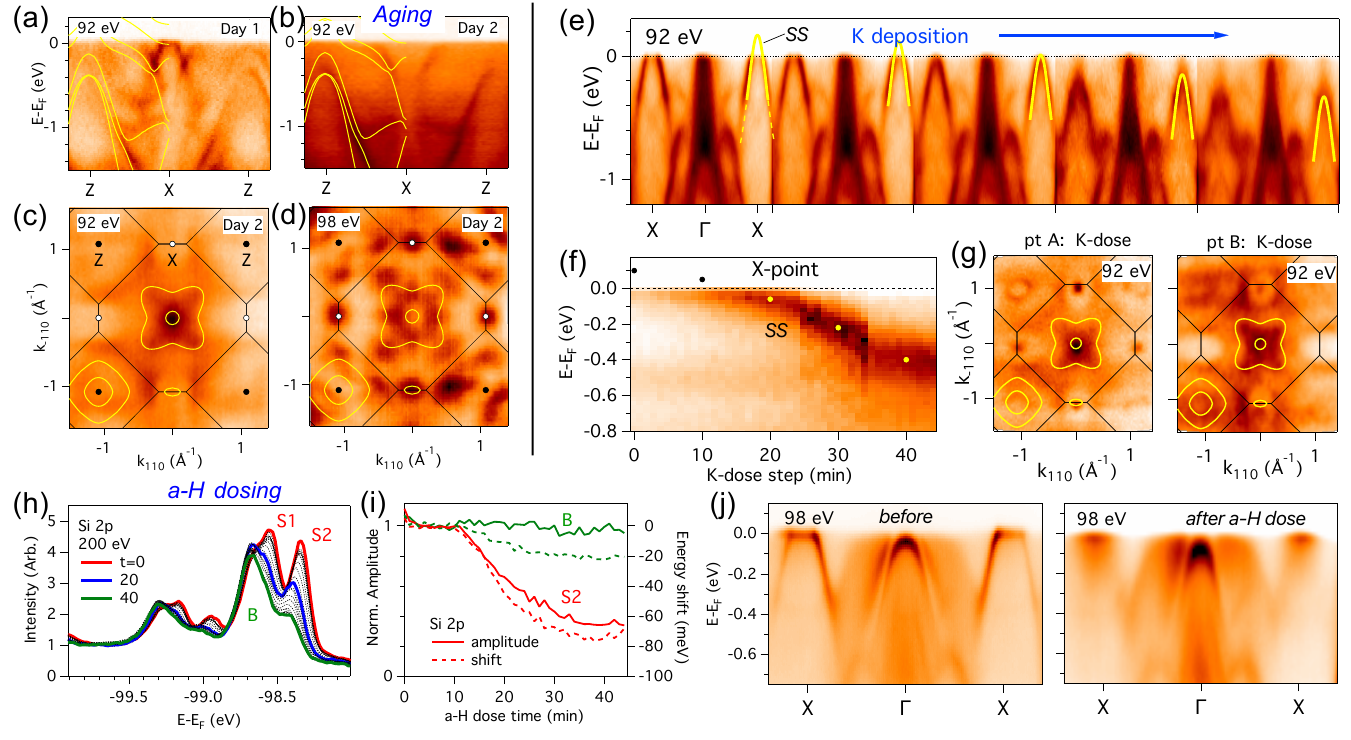}
\caption{
\textbf{Surface modifications of Si-terminated \URS.} 
\textbf{Surface aging:} Z-X-Z VB maps at 92 eV for (a) fresh cleaved and (b) 24 hour aged surface. 
(c) 92 eV and (d) 98 eV FS maps for the aged surface.
\textbf{Alkali deposition:}  
(e) Series of X-\G-X VB maps at 92 eV for select steps of K surface evaporation exhibiting a downwards energy movement of the X-point hole band surface state. The energy-shifted DMFT slab calculation hole-band dispersion is overplotted.
(f) Fine step evolution at the X-point spectrum with K surface deposition.
(g) FS maps at 92 eV after K surface deposition for both A and B surface regions
(h) Fine step evolution of the Si 2$p$ spectra (h$\nu$=204 eV) during the a-H exposure showing the selective suppression of the low-binding energy surface state component. 
(i) Summary analysis of the relative Si 2$p$ bulk and S2 component amplitude attenuation and energy shift from a-H dosing.  
(j) Comparison of X-\G-X VB maps at 98 eV before and after a-H dosing. 
}
\label{dose}
\end{center}
\end{framed}
\end{figure}

\subsubsection{Surface aging} \label{aging}

The approach of isolating the bulk electronic structure by surface modification in hope  of suppressing the \SiT surface states is investigated in this section by three methods of 
(i) surface aging, e.g. slow adsorption of residual gases, 
(ii) surface electron doping via surface deposition of alkali atoms, and 
(iii) surface dosing with atomic hydrogen \cite{Denlinger2010}.
For the first   (natural occurrence) method, figure \ref{dose}(a) and (b) shows the comparison  of a 92 eV Z-X-Z spectral image from a freshly cleaved surface to a second day repeat measurement of the same spot in which the sample temperature was allowed to warm up overnight to $\sim$150 K with an accompanying small increase in the vacuum base pressure before cooling back down to $\sim$20 K.
The initial distinct X-point surface state hole pocket is observed to completely disappear in the aged spectrum, better revealing the day 1 electron-like band dispersion(s) at X which  is actually a hole-like band(s) centered on the Z-point with an apparent larger \kF\ than that of the overplotted DFT band.  
The Fermi-edge intensity along Z-X-Z coming from this band is apparent in the day 2 FS map in figure  \ref{dose}(c) whose features are consistent with the \UT surface FS map of figure \ref{xy}(c).  
Also the aged on-resonance 98 eV FS map exhibits `dots' of high $f$-spectral weight that are even more clear than  for the \UT surface FS in figure \ref{xy}(c), and notably still contains $f$-weight at the X-point despite the absence of the X-point hole-band.
  However, overall, the higher background and diminished contrast of dispersing bands and  the fuzzier FS map, make this not a viable approach for a high quality study of the bulk states. 

\subsubsection{Alkali deposition} \label{alkali}

For the next  (intentional) surface modification method, figure \ref{dose}(e) shows a series of select 92 eV X-\G-X VB maps from a fine sequence of progressive evaporation of potassium atoms onto the Si-terminated region of the cleaved surface. 
The clear effect on the X-point hole-band surface state is to shift it to lower energy until it drops below \EF. 
This is interpreted as the Si $p_z$ dangling bond orbital readily accepting charge donated by the surface K atoms.
A non-parabolic V-shaped tip of the hole band dispersion is revealed when it drops below \EF\ consistent with the DFT+DMFT slab prediction of this surface state in the previous section.
A constant momentum slice of the full K-dosing data set at the center of the X-point hole pocket shows in figure \ref{dose}(f) the progressive movement of the hole band maximum below \EF\ until it saturates at -0.4 eV.    
Again, the transformation of the \SiT cleaved surface 92 eV FS map, shown in figure \ref{dose}(g), is comparable to the \UT surface region, but still with dots of intensity at the X-point possibly coming from the upper Dirac cone of the same Si surface state. 
Also shown is the K-dosed \UT region 92 eV FS map which does not show an improvement over the cleaved FS map, although a complete suppression of X-point intensity in the $k_y$=0 plane is achieved.
Despite the promising bulk-like FS result, the resultant VB map still possesses the strong (energy-shifted) surface state bands and are far from resembling the \UT VB map in figure \ref{xy}(c).

\subsubsection{Atomic-H dosing} \label{hydrogen}

The third surface modification experiment employs exposure of the surface to atomic hydrogen (a-H) in hopes of selectively replacing the Si dangling bond surface state with a Si-H bond. 
Using the Si 2$p$ core-level lineshape to fine monitor the dosing process in figure \ref{dose}(h), indeed the low BE peaks are observed to be preferentially weakened with a-H exposure, and with essentially no reduction in the bulk Si 2$p$ component amplitude as shown in an analysis plot in figure \ref{dose}(h). 
As the Si 2$p$ surface state amplitude is suppressed, the peaks also shift by up to 80 meV to higher BE consistent with an addition of H$^-$ surface charge. The bulk Si 2$p$ component also shifts by a smaller amount of 20 meV. 
A comparison of on-resonance 98 eV valence band X-\G-X VB maps for before and after a-H dosing, shows the near disappearance of the X-point hole band intensity,  but still with distinct \EF\ $f$-weight at the X-point similar to the aged 98 eV FS map in figure \ref{dose}(d).  The reason for this is explored in Section \ref{xpt}.
Also the 30 meV hole surface state at \G\ is weakened but not suppressed due to its second layer Ru-d character origin that is less  influenced by the creation of Si-H bonds. 
The utility of this a-H dosing approach, however, is to further suppress the already weak residual Si surface state intensities in the \UT regions.

\section{Electronic structure mapping below 200 eV} \label{arpes200}

\subsection{ThRu$_2$Si$_2$ $f^0$ reference} \label{trs200}

\sectionmark{FIG. 5. TRS 150eV }
\begin{figure}[t]
\begin{framed}
\begin{center}
\includegraphics[width=14cm]{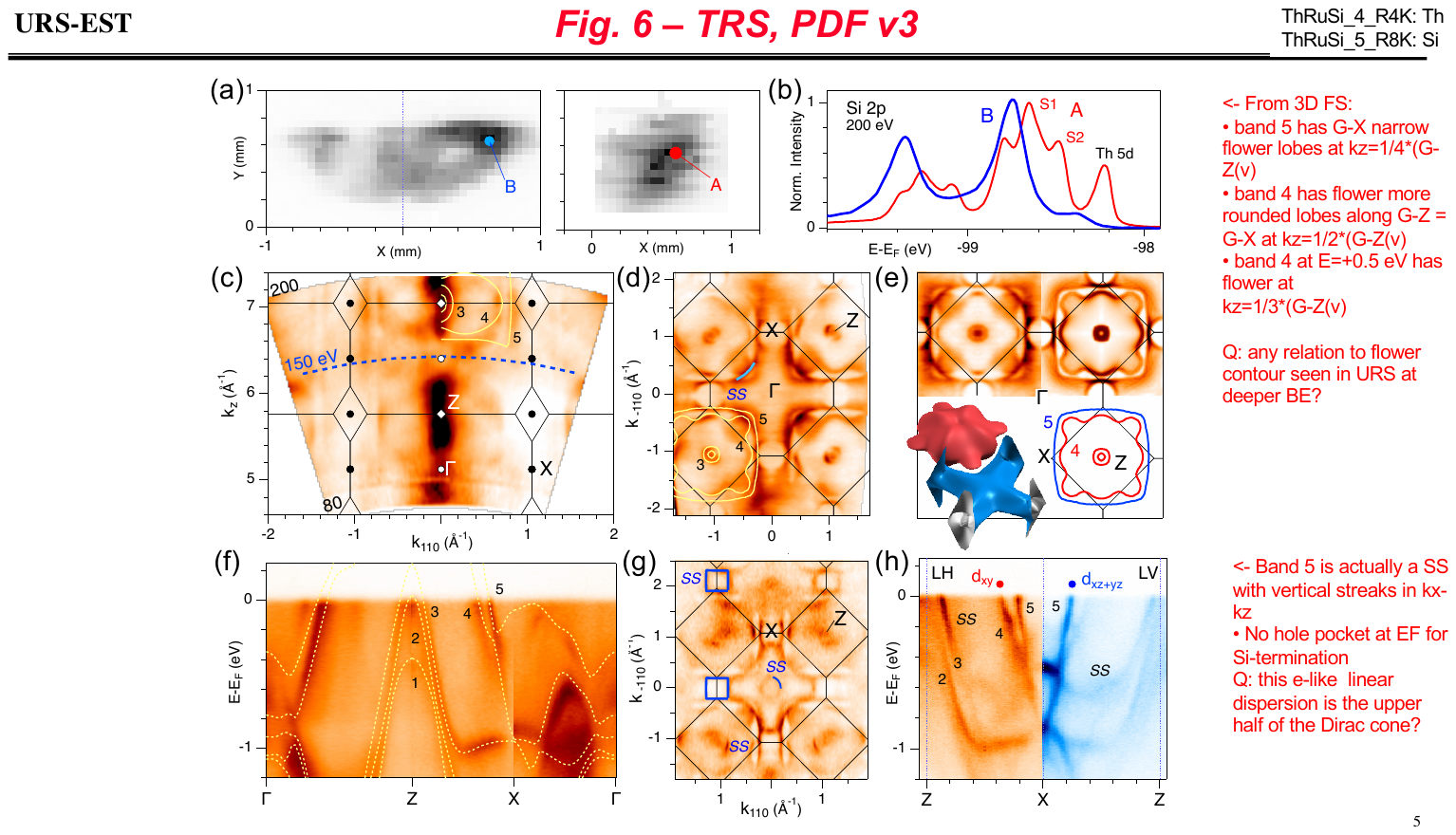}
\caption{
\textbf{Electronic structure of $f^0$ reference \TRS.}
(a) Spatial maps of two sub-mm sized \TRS\ crystals that separately generated \SiT(A) and \ThT (B) ARPES data.
(b) Comparison of the Si 2$p$ spectra from the optimized spots of the two crystals. 
An additional Th 4$d_{3/2}$ peak also appears in the low BE region.
(c) Normal emission photon dependent FS map of the \ThT\ sample in the \abdir-\cdir\  plane spanning 80-200 eV.
(d) High symmetry \kxky\ Fermi surface map at hv=150 eV.
Overplotted DFT FS contours provided in (c,d) showing excellent agreement.
(e) ARPES FS map four-fold symmetrized about the Z-point (upper left) compared to a \kz-broadening simulation of the \G-plane FS spectral intensities (upper right). (Lower) DFT band 4 and 5 2D FS contours and 3D FS sheets.  
(f) High symmetry band dispersion cuts along \G-Z-X-\G\ at 150 eV with overplotted DFT bands.
(g) FS map of the \SiT\ surface of \TRS\ with identification of surface states at \G, Z and X. 
(h) X-Z bands of \SiT\ \TRS\ highlighting the strong LH/LV polarization dependence of the large \kF\ hole bands (4,5) with in-plane and out-of-plane $d$-orbitals, but also identifying additional surface state band features.
}
\label{trs}
\end{center}
\end{framed}
\end{figure}

Before further presentation of the quest for the bulk electronic structure of \URS, we first present the experimental ARPES electronic structure of the $f^0$ reference compound  \TRS\ in comparison to theoretical DFT calcuations.
This is a useful starting point for introducing our indexing of the non-$f$ bands, and our methods for theory comparison (where the agreement is excellent). This non-$f$ band reference also prepares the reader to understand the near \EF\ \URS\  E($k$) band dispersion plots presented further below.

Cleaved crystals of \TRS\ inherently have the same surface termination issues as \URS, except that in the measurements here, only tiny sub-millimeter size crystals without polycrystalline orientational domains were available. Hence the examples of Si- and \ThT ARPES data come from two separate crystals whose spatial mapping is shown in figure \ref{trs}(a).  Also the \EF-intensity contrast mechanism of the \dtof\ resonance is inherently not available for Th whose unoccupied 5$f$ states are $\sim$1 eV above \EF,  and so the Si 2$p$ spectra at the optimized spot(s) provided the primary tool for identifying the surface termination.  
Similar to what occurs for \URS, the Si 2$p$ spectra shown in figure \ref{trs}(b), exhibit the same distinct characteristic multi-peaked lineshape signature of the two \SiT low BE surface states in contrast to the simpler lineshape for Th-termination.  
First we focus on the bulk-like \ThT results of photon dependent \kxkz FS mapping, the 150 eV \kxky\ FS and band dispersions before comparing  to the \SiT results.

In figure \ref{trs}(c) the normal emission \abdir-\cdir\  plane \kxkz FS map shows the clear signatures of a large \kF\ FS contour centered on both Z-points. This FS contour has good size and shape correspondence to one of the overlaid DFT bulk FS contours (band 4). 
The \kxkz FS map also exhibits some vertical streaks around the -X-X-X- zone boundary that are typically indicative of surface state character. 
 However, in this case, an even larger bulk DFT contour (band 5) also contains near vertical edges,  thus providing an alternative \kz-broadened bulk origin explanation for the measured vertical streaks. 
For the high symmetry 150 eV FS map  of figure \ref{trs}(d), the \G\ and X-points are very clean of any \EF\ intensity and all the features come from the second BZ Z-points including (i) a small circular \kF\ pocket surrounding a dot of intensity at Z, (ii) nicely sharp and continuous FS contours of a large diamond-shaped contour just inside the BZ boundary with rounded tabs pointing towards \G, and (iii) an even larger surrounding squarish contour.  
All of these Z-point shapes and sizes,  including the finding of two small \kF\ contours, is nicely reproduced by the  DFT FS contours overplotted in one quadrant.

Another method for experiment-theory comparison is shown in the upper quadrants of figure \ref{trs}(e) where the experimental FS map has been four-fold symmetrized about the Z-point and the DFT FS has been converted into a DOS intensity image with a partial averaging of the three-dimensional (3D) FS (lower left view) along $k_c$ using a Lorentzian 0.2 \invA\ weighting.  
This simulation of experimental \kz-broadening effects reproduces (i) the enhancement of intensity for vertical edges of the 3D FS such as the vertical circular cylinder of  the band 5 sheet at the X-point, and (ii) a relatively weakened intensity for more dispersive features such as the thin ``tabs'' of the 8-lobe flower band 4 sheet whose  3D shape is  also illustrated in a 3D view in figure \ref{trs}(e) inset. 

Figure \ref{trs}(f) shows the combined \G-Z, Z-X and X-\G\ band dispersion cuts from the 150 eV map data with overlay comparsion of the \TRS\ DFT bands indexed as 1-5.  Again there is quite good agreement with theory including evidence of the two very small hole pocket \EF\ crossings at Z and the two large \kF\ hole-like bands with  large (light mass) 2.5-4.0 eV-\AA\ band velocities that come together at $\approx$0.3  below \EF. 
The $f$-$c$ hybridized dispersions of bands 4 and 5 along Z-X and \G-Z are key points of discussion for \URS.

With the bulk electronic structure showing excellent agreement to DFT, we now return to the \SiT FS in figure \ref{trs}(g) for comparison. 
Here we can identify a clear new circular surface state contour at \G, consistent with the \TRS\ slab calculation, and now a 4-fold symmetric square FS contour at X.  Also a sharp larger-\kF\ contour observed surrounding the Z-point intensity is yet another surface feature.
Exploring the \SiT Z-X cut in figure \ref{trs}(h), we find even sharper bands than for the \ThT surface and some strong LH versus LV polarization asymmetries.  The large-\kF\ bands 4 and 5  dispersions are observed more clearly to extend down to -0.4 eV before crossing. While band 5, with theoretical out-of-plane  d$_{xz+yz}$ orbital character, is observed to extend down to -0.8 eV for LV polarization, the in-plane d$_{xy}$ orbital character band 4 is observed to be highly suppressed for (in-plane) LV polarization of the incident x-rays.
In addition we observe the Z-point surface state dispersion as well as a lower BE energy shifted replica of the flattish -1 eV band along X-Z.

\subsection{\URS\ comparison to DFT and DFT+DMFT} \label{urs200}
 
\begin{figure}[t]
\begin{framed}
\begin{center}
\includegraphics[width=15cm]{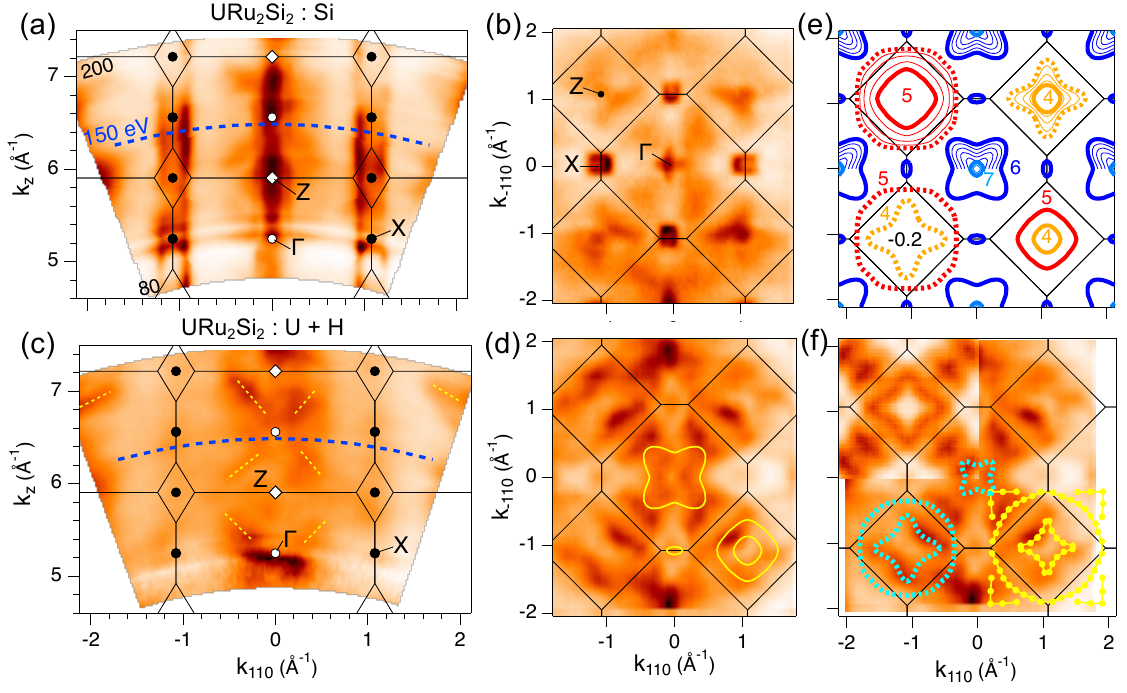}
\caption{
\textbf{Fermi surface mapping of \URS\ below 200 eV.}
(a) Normal emission photon dependent FS map of \SiT \URS\ in the \abdir-\cdir\ plane exhibiting strong \kz-independent surface state features.
(b) Corresponding high symmetry FS map at 150 eV. 
(c) Normal emission \kxkz\ FS map \UT \URS\ with additional atomic-H dosing that is absent of vertical \kz\ streaks.
(d) Corresponding FS map at \hv=150 eV.
(e) Theoretical $f$-itinerant DFT \G-plane \EF\ contours (thick solid lines) and higher BE contours (thin, dashed lines) of \URS: 
(center+lower right) \EF\ contours with two squarish Z-point hole pockets, 
(upper) hole band 4 and 5 contours for binding energies down to -0.2 eV showing transformation to larger flower and circular contours, 
(lower left) combined bands 4 and 5 contours for -0.2 eV.
(f) ARPES FS map with four-fold  symmetrization about the Z-point (upper left) with overlays of -0.2 eV DFT contours (lower left) and SX-ARPES derived FS contours \cite{Kawasaki2011} (lower right).
}
\label{urs150}
\end{center}
\end{framed}
\end{figure}

We further our comparison between \SiT and \UT regions  to wider photon dependent mapping 
up to 200 eV, where we include the a-H dosing of the \UT regions and select a new \G-plane cut at 150 eV for analysis and theory comparison where the $f$-states are not resonantly enhanced.
First in figure \ref{urs150}(a) and (c),  respectively, we plot the normal emission \abdir-\cdir\ plane \kxkz FS maps of  \SiT \URS\ and atomic-H dosed \UT \URS\ which span two BZs along $k_c$.

\sectionmark{FIG. 6. hv, FS 150eV }

While the \kxkz FS map of \SiT \URS\ is dominated by \kz-independent surface state vertical streaks along -X-X-X-,  and a  $k_x$=0 strong intensity vertical streak coming from the \Gbar\ surface states, such vertical streakiness is  completely absent in the \UT FS map.
Instead the highest intensity features are somewhat broad diagonal structures whose \kz-dispersion is a signature of 3D bulk character. 
These diagonal bulk features are also  observed with weaker relative intensity in the \SiT FS map.

The overplotted Brillouin zones in the  \kxkz FS map plots identify 150 eV as another approximate high symmetry \G-plane photon energy cut that is then compared in figure \ref{urs150}(b) and (d). 
Similar to the 92 and 98 eV \G-plane FS maps in figure \ref{xy}, the \SiT FS map exhibits the strong intensity squarish 4-fold symmetric X-point surface states and a high intensity at the central \G-point.
For the \UT 150 eV FS map, the \G-point is now a local intensity minimum and the X-point only has a faint residual \EF\ intensity as noted earlier for 98 eV. Again, the maximum intensity features, which are also observable in the \SiT FS map with weaker relative intensity, are somewhat broad and generally patchy making it hard to \textit{a priori} identify continuous FS contours. 
 However,  after  4-fold symmetrization about a second BZ Z-point, performed in the upper left quadrant of figure \ref{urs150}(f), one can postulate a star-like contour with lobes pointing towards the X-points.

While the DFT FS contours showed some agreements to the 92 eV FS maps, they again, similar to  what occurs for 98 eV, do not show a correspondence here at 150 eV especially for the Z-point squarish theory contours compared to the flower-like experimental shape. 
To explore this discrepancy, we examine the evolution of the DFT calculation constant energy contours below \EF\  in figure \ref{urs150}(e) in 50 meV steps down to -0.2 eV with different color-coding of each band index.  
The  \G-point DFT flower contour (band 6) is filled in towards the center, identifying its shallow electron pocket origin.
At the Z-point the bands 4 and 5 energy contour evolutions are separately explored in the upper quadrants. The larger squarish \EF\ contour (band 5) evolves outwards (hole-like) to a larger more circular shape, while the smaller \EF\ square (band 4) evolves to the star-like shape that is observed experimentally. 
The most rapid changes of wave vector versus binding energy highlight the heaviest effective mass band dispersion regions. While the band 5 FS is isotropically heavy in all directions, the band 4 and band 7 FS sheets are  especially heavy in the direction of the lobes.

The -0.2 eV contours are then plotted together in the lower left quadrant of figure \ref{urs150}(e) and also onto the experimental FS intensity map in figure \ref{urs150}(f).
Finally in the lower right quadrant, a comparison is made to the FS contours derived from the bulk-sensitive 760 eV SX-ARPES measurements by Kawasaki \etal\ \cite{Kawasaki2011}. 
Both the star-like band 4 contour and the large band 5 circular contours are consistent with the \EF-tuned DFT theory contours, albeit with a  slightly smaller star (larger circular) size.  In addition, the experimental 150 eV intensities along \G-Z in the first BZ are derived at 760 eV  to be triangular features bordered  on one side by the band 5 circular contour.

\sectionmark{FIG. 7. VB 150eV }
\begin{figure}[b]
\begin{framed}
\begin{center}
\includegraphics[width=15cm]{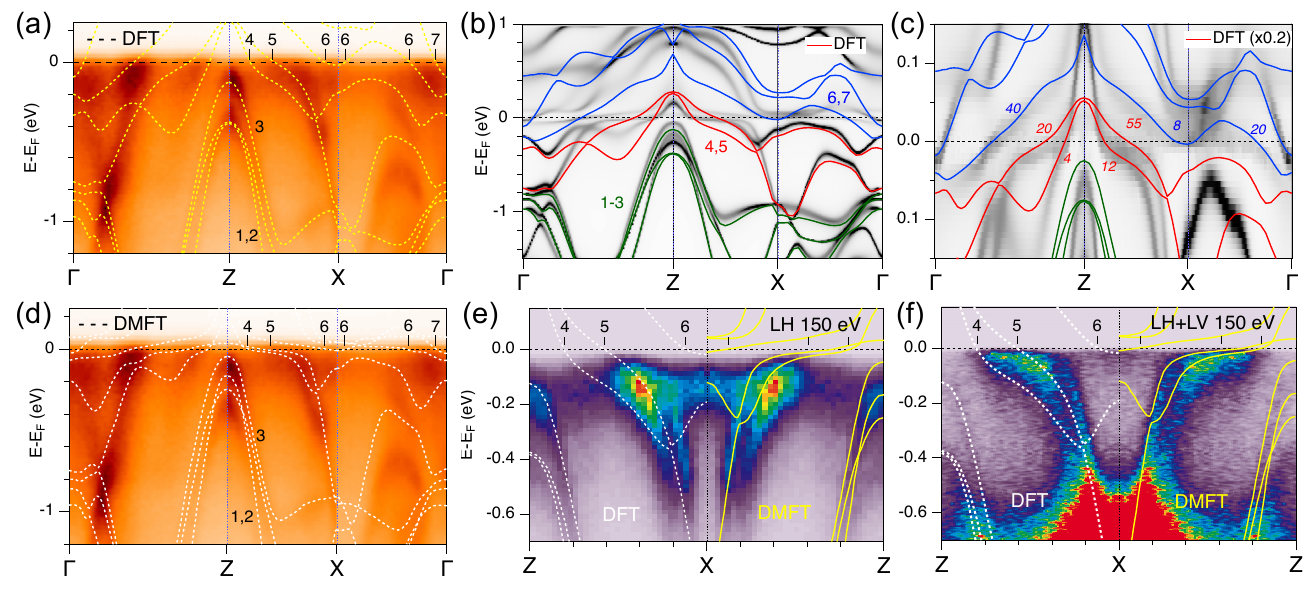}
\caption{
\textbf{ARPES high symmetry band dispersion comparison to DFT and DFT+DMFT calculations.}  
(a) ARPES \G-Z-X-\G\ band dispersion cuts of \URS\ measured at 150 eV with overlay comparison to $f$-itinerant DFT band structure. 
(b) DFT+DMFT \Akw\ spectral function with comparison to DFT bands. 
(c ) Near \EF\ comparison of the DFT+DMFT calculation to a 5$\times$ energy renormalization of the DFT bands.
DMFT effective masses (m*/m$_e$=7.62 \kF/\vF) are labeled at various \EF\ crossings.
(d) ARPES \G-Z-X-\G\ band dispersions compared to DFT+DMFT bands.
(e,f) Near \EF\ comparison of ARPES Z-X-Z to both DFT and DFT+DMFT bands for (e) LH polarization (with geometry \#1, figure \ref{intro}(f)) and (f) LH+LV polarization (with geometry \#2,  figure \ref{intro}(f)) and higher energy resolution. 
}
\label{vb150}
\end{center}
\end{framed}
\end{figure}

To explore  further this discrepancy between FS contours and band theory, we extract and combine the 150 eV high symmetry band dispersion cuts along \G-Z, Z-X, and X-\G\ in figure \ref{vb150}(a) and make an overlay comparison to the $f$-itinerant DFT bands.  
First we note the correspondence of the \EF\ weight in the band plot compared to the bands 4 and 5 \EF\ weight, size and shape.  Along \G-Z there are two locations of high \EF\ weight with large separation, one closer to \G\ (band 5) and the other closer to Z (band 4). 
In contrast, along Z-X there is one broad location of high \EF\ weight where two closely spaced bands are resolved at higher binding energy close to X.
This corresponds to where the band 4 flower lobe \kF\ becomes maximum and comes close to band 5.
These two bands were cleanly resolved in \TRS\  linearly dispersing from -0.4 eV  to \EF.  Here they exhibit the expected $f$-$c$ hybridization dispersion rollover towards heavier effective mass as soon they enter the strong $f$-peak region around -0.1 eV. 

Similar to \TRS, decent agreement is found with the (overplotted dashed line) DFT bands at high binding energy below -0.2 eV. Closer to \EF, specific notable differences occur. Along X-Z the DFT bands 4 and 5 cross at a shallower energy of -0.3 eV and are immediately curving towards Z where they have small values of \kF\ corresponding to  the two squarish DFT FS contours at Z. 
The result is that the DFT band 5 has a similar \EF\ correspondence to the experimental band 4 high \EF\ weight.
The discrepancy is also severe along \G-Z, where band 5 experimentally looks to disperse from below -1 eV up to \EF\ closer to  \G\ than to Z. However the DFT predicts this band to rollover and cross \EF\ closer to Z and band 4.   

The two natural concerns  with this comparison are (i) the well-known lack of strong DFT electron correlations that induce energy renormalization of the $f$-states and (ii) lack of experimental resolution to  resolve very low energy scale dispersions of the renormalized DFT bands near \EF.  
To address (i) we introduce for comparison a dynamical mean field theory (DFT+DMFT) calculation, whose 20 K spectral function \Akw\ image is shown in figure \ref{vb150}(b) with comparison to the DFT bands.  
As expected there is agreement at higher BE, but closer to \EF\ there is  in figure \ref{vb150}(b) a moderate narrowing of the $f$-$c$ hybridization energy scale.  
A zoom in of this comparison \ref{vb150}(c) with the DFT band energies rescaled by 0.2, shows an overall match of bands within 50 meV of \EF, and also with no change in the \EF\ crossing momentum \kF\ values.  
Hence this DFT+DMFT calculation introduces no new predictions of different FS sizes and shapes compared to DFT, but makes much more visually apparent the  small band velocities $<$0.1 eV-\AA\ and large effective masses, up to m*=55m$_e$ as  labeled for various \EF\ crossings.   
A similar $>$1 eV wide band dispersion comparison of the ARPES to the DFT+DMFT bands with 5$\times$ $f$-state renormalization in \ref{vb150}(d) is much more overall satisfactory.

A narrower energy range zoom comparison to ARPES Z-X-Z  in figure \ref{vb150}(e) highlights in more detail the  \kF\ discrepancies for the DFT and DFT+DMFT predictions.  To test concern (ii), later higher resolution measurements were performed in the second  ARPES  geometry (\#2) with a vertical slit and with variable x-ray polarization.   In figure \ref{vb150}(f), LV polarization assisted by  better resolution ($<$20 meV) is observed to transform the highest spectral weight from the 0.1 eV turning point of the band 4 and 5 dispersions (closer to X) to the tip of very heavy band dispersions in agreement to the predicted band 4 \kF\ close to Z.  
Also the sharp corner of the spectral intensity of the electron side of band 5 as it reaches \EF, indicates that the 5$\times$ renormalization narrowing of $f$-band velocities near \EF\ of this DMFT calculation is a significant improvement. 
As discussed in section \ref{zonefold}, the true renormalization factor may be as large as  20$\times$ narrower than the DFT prediction from comparison to a high-resolution $M$-band feature \cite{Bareille2014}. 
The \EF\ spectral weight at the X-point and its consistency with a shallow bulk DFT electron pocket is discussed further in section \ref{xpt}.

Hence we must conclude that the DFT-predicted FS cannot be  reliably assessed based on measurements with  inadequate energy resolution, especially for such extremely heavy mass bands.  
The discrepancy with the $f$-weight distribution in the ARPES mapping likely arises from the  small band velocities giving $f$-weight  that resides just below \EF\ and extends over a large range of momentum.  
Another potential factor is that the $T$=0 ground state FS is not expected to be fully achieved in the PM phase just above the HO phase. 
In addition, specifics of the $f$-orbital anisotropies and photoionization matrix elements for different incident x-ray polarizations and experimental geometries may play a role.
We explore this possibility further at higher photon energy in the next section.

\section{Soft X-ray ARPES} \label{sxarpes}

Another standard approach for isolating the bulk electronic structure is to increase the photoelectron probe depth by the use of higher photon energy excitation.  
In this section, we present some example ARPES mapping in the 180-500 eV range (Region IV) which is in the lower range of the so-called ``SX-ARPES'' regime, where the photoelectron probe depth is $\sim$10 \AA, and we make comparison to the bulk sensitive ARPES results of Kawasaki \etal\ \cite{Kawasaki2011} from Region V (680-760 eV) where the probe depth may exceed 15 \AA.  

\sectionmark{FIG. 8. SX-ARPES } 
\begin{figure}[t]
\begin{framed}
\begin{center}
\includegraphics[width=15cm]{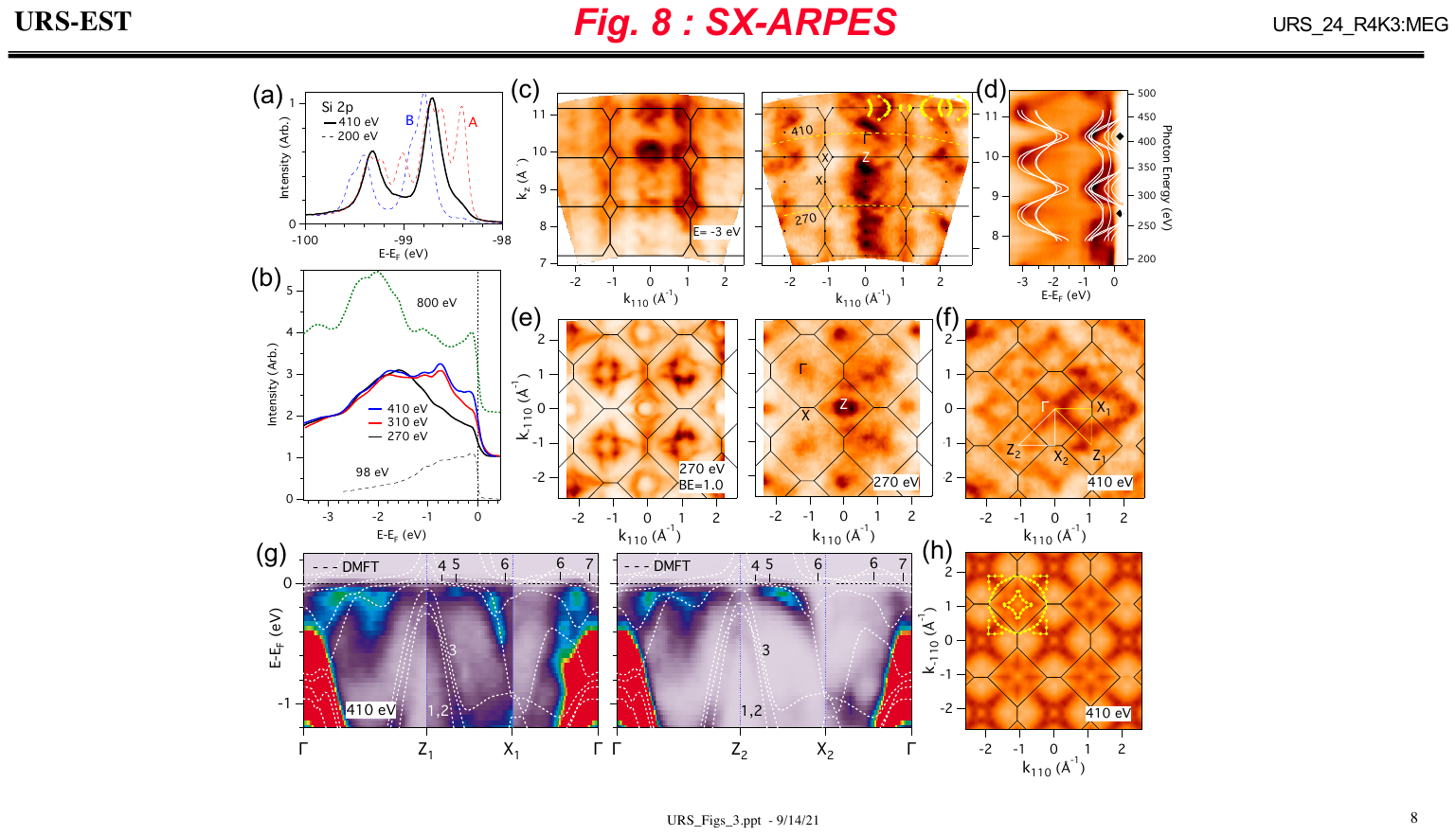}
\caption{
\textbf{High energy ARPES band mapping of \URS\ above 200 eV.}
(a) Si 2$p$ core-level spectrum measured at 410 eV compared to low energy spectra (dashed).
(b) Angle-integrated valence band spectra for 270, 310 and 410 eV compared to an 800 eV spectrum (from \cite{Fujimori2016}) and to the 98 eV resonance spectrum. 
(c) Normal emission photon dependent maps in the diagonal \abdir-\cdir\  plane for -3 eV binding energy and for \EF, 
with overplot of  $>$680 eV SX-ARPES-derived FS contours \cite{Kawasaki2011}. 
(d) Normal emission VB map showing periodic dispersion along -Z-\G-Z-\G-. 
(e) High symmetry Z-centered $k_x$-$k_y$ map at 270 eV for BE=-1 eV and \EF.
(f) High symmetry \G-centered \EF\ $k_x$-$k_y$ map at 410 eV.
(g) High symmetry band dispersions along \G-Z-X-\G\ for the two different 410 eV $k$-paths marked in (f) with a comparison to DFT+DMFT bands.
(h) Symmetrized 410 eV FS map with overplot of 760 eV derived FS contours \cite{Kawasaki2011}.
}
\label{sx}
\end{center}
\end{framed}
\end{figure}

First in figure \ref{sx}(a) we make note of the much simplified Si 2$p$ core-level lineshape measured at 410 eV (KE$\sim$310 eV) compared to the 200 eV (KE$\sim$100 eV) spectra from the different surface terminations. It exhibits a single bulk component with a weak low BE surface component owing to the increased bulk sensitivity. 
Next we compare the angle-integrated valence band spectra of \URS\ (averaged from entire angle maps) at three photon energies in this 200-500 eV region in figure \ref{sx}(b).
The progressive increase in weight near \EF\ at higher photon energy reflects the U 5$f$ photoionization cross section rising from a Cooper minimum at 200 eV.   
At even higher photon energy, the 800 eV spectrum from Fujimori \cite{Fujimori2016} becomes even peakier close to \EF, and can be claimed to be the dominant character at \EF\ in that higher energy range.
Also compared is the averaged valence spectrum from a 98 eV map from the \UT surface, which shows an extreme dominance of U 5$f$ weight arising both from the \dtof\ resonance and from the presence of surface U atoms, as well as from the Ru 4$d$ Cooper minimum.

A normal emission photon-dependent map of \URS\ measured in the \abdir-\cdir\  plane, shown in figure \ref{sx}(c), compares a 3 eV binding energy slice to the \EF\ intensities. The high BE map slice exhibits quite clear periodicity of the band structure for easy determination of the crystal inner potential parameter ($\sim$14 eV) used in the $k$-space conversion, and good \kz-resolution that easily distinguishes the different 90$^{\circ}$-rotated X-points at the BZ boundary. 
In contrast, upon tuning the energy within 0.5 eV of \EF\ where a broad band of $f$-weight exists, the spectral intensities become quite complex and with less symmetry, indicating an enhanced sensitivity of the initial state $f$-orbitals to the polarization geometry and matrix element effects as compared to the non-$f$ orbitals at deeper BE.   This is a common theme for ARPES measurements of \URS\ experienced at most photon energies, and is opposite to ARPES of many other systems in which the band structure dramatically sharpens up at \EF\ where the quasiparticle (QP) lifetimes are greatest.
Nevertheless, at higher photon energies above 350 eV, a repetition of large oval shaped structures centered at the Z-points is visible in the top three BZs.
Bulk FS contours derived from the $>$680 eV SX-ARPES of \URS\ by Kawasaki \etal\ \cite{ Kawasaki2011} are overplotted and show good consistency with the outermost contour. 
The band dispersion -\G-Z-\G-Z- periodicity along \kz\ identifying electron-like (hole-like) \G\ (Z) points is also nicely observed in the  normal emission VB map in figure \ref{sx}(d), where the clarity of the -1 eV and -3 eV band periodicity similarly improves at higher photon energy and increasing probe depth.  

While angle-dependent FS maps are generally more interpretable than photon dependent maps due to greater constancy of matrix elements, this is not the case for the two FS maps shown in figure \ref{sx}(e) and \ref{sx}(f) at selected photon energies of 270 eV and 410 eV that cut through normal emission Z- and \G-points, respectively.
Again the higher binding energy -1 eV slice shows much greater clarity of distinguishing the different \G\ and Z-point contours with uniform and symmetric intensities, while the two FS maps, although showing correlation to the overplotted BZ, exhibit strong intensity asymmetries about $k_x$=0 reflecting a greater sensitivity to the broken symmetry of the experimental geometry of the incident angle of x-rays linearly polarized in the $k_y$=0 plane.

We further explore this intensity asymmetry in two different  \G-Z-X-\G\ band structure cuts in figure \ref{sx}(g) taken from the 410 eV map data set with comparison to DFT+DMFT bands introduced in the previous section.  
The high intensity convergence of bulk bands at -0.8 eV at \G\ is saturated in the color table intensity scale to reveal general agreement with the theory bands closer to \EF\ along \G-Z and Z-X. 
While the \G-Z$_1$ and \G-Z$_2$ cuts are consistent with each other in showing a midway shallow electron dispersion involving bands 4-6, the two different Z-X cuts exhibit a stronger selectivity of band intensities, that is reminiscent of the 150 eV ARPES behavior in figure \ref{vb150}(e) and (f). 
While the Z$_2$-X$_2$ cut shows an extended hole-like intensity that could be due to both bands 4 and 5, the Z$_2$-X$_2$ cut highlights the light hole band dispersion down to -0.6 eV close to X, and shows a suppression of the \EF\ weight, but still retains a high intensity dot of \EF\ weight close to the theory band 5 \EF\ crossing. 
Along \G-X there are a few notable theory-to-experiment discrepancies, also consistent with 150 eV ARPES observations,  including a too shallow energy of theory band 4 and a high intensity spectral weight  close to \G\ in between bands 5 and 6.

Finally in figure \ref{sx}(h), the 410 eV map has been four-fold symmetrized about the center \G\ point and also symmetrized to the outer BZs. 
 This reduces the description of the FS data set to two main intensity features of radial segments of high \EF\ intensity along \G-Z 
 that result from the heavy mass bands 5 and 6, and along Z-X resulting from the heavy mass bands 4 and 5.
In addition, there is a weak intensity dot at X still visible consistent with the lower photon energy result previously presented.
The SX-ARPES derived 760 eV  FS contours from Kawasaki \etal  \cite{Kawasaki2011} are also overplotted onto this symmetrized FS map. 
The larger circular FS contour of band 5 is not visible in this 410 eV FS map.
Similar anisotropic in-plane intensity enhancements along [100] around the \G\ point and along [110] around the Z point have also been identified by Fujimori \etal\ \cite{Fujimori2021} from an SX-ARPES mapping of the full 3D BZ of \URS\ in comparison to DFT-based simulations of the ARPES intensities.   

Our conclusion from this SX-ARPES section is that indeed, while signatures of the bulk band structure at higher binding energy are more clearly prominent and show suppression of surface states identified earlier, the observed FS patterns of intensity derive from $f$-weight of very heavy mass bands over extended lines of momentum that are too energy resolution limited to enable quantification of the detailed \EF\  crossing.

\section{Temperature dependence of bulk and surface states} \label{tdep}

In this section, global \kxkz FS mapping is performed with variable x-ray polarization control in region II, with attention to the lower energy range \textit{below} the 98 eV U 5$f$ resonance condition where the experimental energy resolution is improved even further, thus allowing meaningful $T$-dependent measurements into the HO phase. 
First, we revisit the \cdir\ \Gbar\ region along normal emission to map and help clarify the \G- and Z-point zone folding that generates the $M$-band feature in the high-resolution ARPES literature, and discuss $T$-dependences experimentally observed there.
Second,  we  revisit the high symmetry zone boundary X-point and clarify the surface versus bulk character of the shallow electron-band U 5$f$ state relative to the hole-band Si $p_z$ surface states. 
Finally, attention is focused on \EF\ states near the BZ boundary along \adir\  which are seemingly incongruent with the DFT Fermi surface prediction.   
This leads to a theoretical re-examination of the N-point where another region of heavy mass bands is revealed, and new alternate origins of commensurate and incommensurate scattering vectors are discussed.
The temperature dependence of the heavy band mass states in this region reveals evidence of a Kondo hybridization evolution beginning at high $T$ that is interrupted just above \THO.

\subsection{\texorpdfstring{$\Gamma$}{G} and Z-points} \label{gzpt}

\sectionmark{FIG. 9. G SS }
\begin{figure}[t]
\begin{framed}
\begin{center}
\includegraphics[width=15cm]{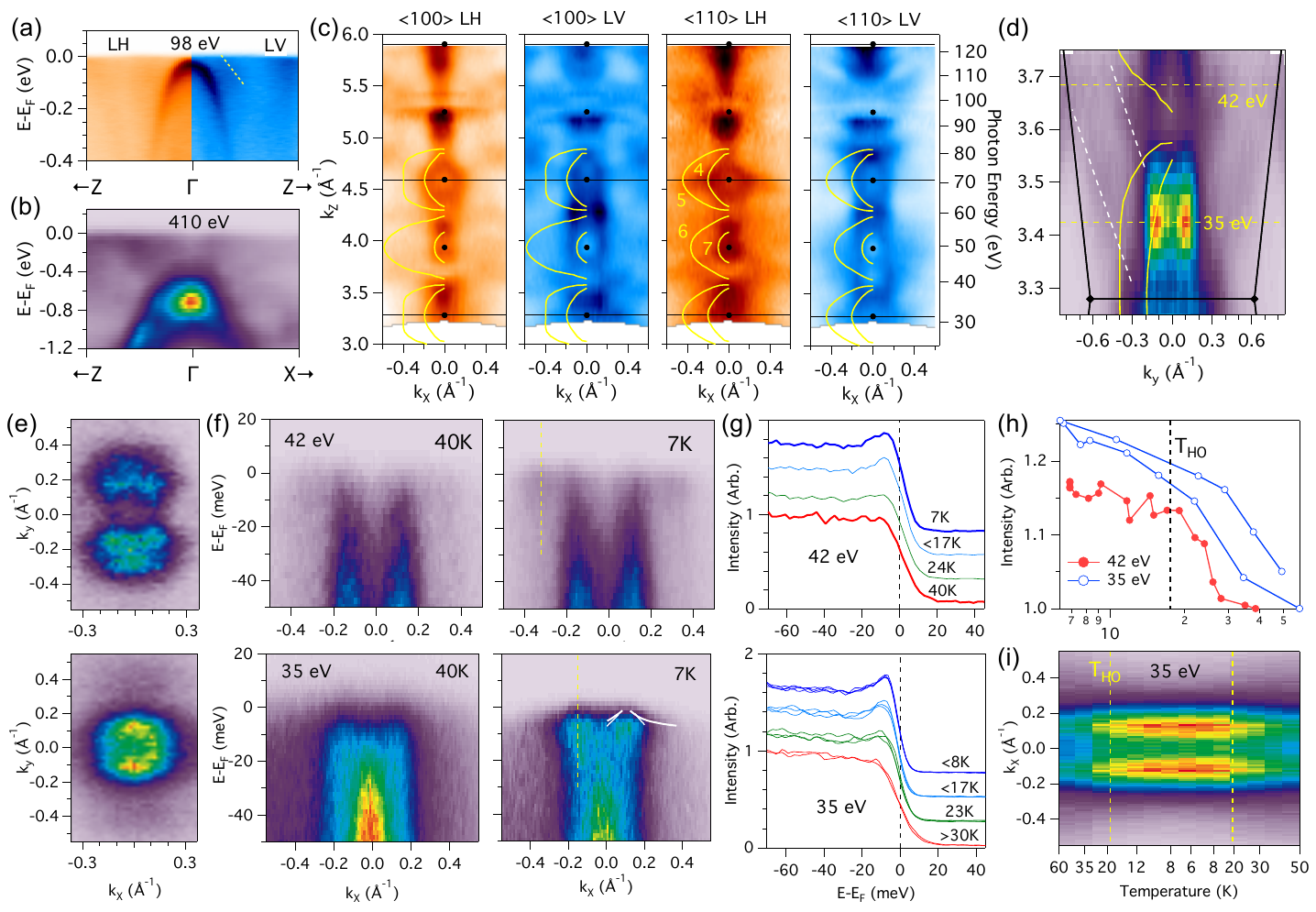}
\caption{
\textbf{\Gbar\ surface and bulk states.} 
(a) Normal emission zoom of the 98 eV \G-point VB map for the Si-terminated surface
(b) Normal emission zoom of the \G-point VB map at bulk-sensitive 410 eV.
(c) Normal emission photon-dependent FS maps for LH and LV polarization and two \adir\ and \abdir\ orientations.
(d) Fine normal emission FS map for \adir\ LV polarization. 
Temperature dependent analysis at 42 eV and 35 eV including 
(e) \kxky\ FS maps and (f) VB maps at 40 K and 7 K, (g) line spectra at the indicated locations in (f), and  
(h) $T$-dependent profiles of the narrow band \EF-weights.
(i) \EF-weight image for 35 eV for cooling and warming through \THO.
Experimental $M$-band dispersions at 31 eV \cite{Bareille2014} and 34 eV \cite{Yoshida2013} are overplotted onto the 35 eV spectrum in (f).
}
\label{gss}
\end{center}
\end{framed}
\end{figure}

First, we zoom into the normal emission region of the on-resonance 98 eV VB map from figure \ref{xy} for the Si-terminated surface (point A) 
where the key features highlighted in figure \ref{gss}(a) are an intense hole-band surface state with strong $f$-weight and -40 meV maximum energy, and a much weaker outer hole-band crossing at \kF=$\pm$0.15 \invA.  
Both states feature prominently in the high resolution studies at lower 7-34 eV photon energies which additionally reveal a narrow  $M$-shaped dispersion within 10 meV of \EF.   
While the DFT+DMFT surface slab calculation in figure \ref{slab}(g) does predict a similar $M$-shaped dispersion at \Gbar,
such a surface-state assignment seems contrary to the experimental observation that the narrow $M$-shaped bands exist only in the hidden order state  \textit{below} 17.5 K,  
which  is rationalized as a bulk zone-folding effect below \THO.
Also provided is a zoomed reminder in figure \ref{gss}(b) of the alternative bulk \G-point electron-like band structure from the bulk-sensitive 410 eV VB map from Section \ref{sxarpes} which instead shows an electron-like dispersion to \EF\ above a -0.7 eV bulk band high symmetry convergence at \G.

To investigate this further, we first provide a more global view context of the normal emission region in figure \ref{gss}(c) for both LH and LV x-ray polarizations and for two \adir\ and \abdir\ orientations of the surface.  
While much of these normal emission features are implicated as having surface state origin from their absence in the \UT FS mapping in figure \ref{urs150}(c), it is clear that there is quite a bit of variations of features along \kz\ that correlate to the bulk Brillouin zone and suggest either coexistence with or coupling to bulk states.
In contrast to the \Mbar\ surface states which appear as extended vertical \kz-streaks in \ref{urs150}(a), because they exist in a bulk-projected gap along -X-X-X-, the \Gbar\ surface states must navigate through the large and small Z-point hole FS sheets. 

Next we zoom in on the normal emission FS map in figure \ref{gss}(d) for \adir\ and LV polarization, i.e. close to the conditions found in the literature, and explore the $T$-dependent spectra at two photon energies of 35 eV and 42 eV which exhibit distinctly different \EF\ fine structure. 
The photon dependent map exhibits a strong intensity enhancement at 35 eV and weak \EF\ intensity features with upwards diagonal \kz-dispersion that appear to emerge from the central region just below and above the high intensity region. 
The upwards \kz-dispersion towards the 50 eV bulk \G-point is consistent with the high-resolution observation claims observing a shallow \textit{electron} band that extends outside the central hole-band region.   
The map here provides the context that the observed shallow electron band is part of a larger \kF\ structure around the \G-point.
At 35 eV the angular map in figure \ref{gss}(e) reveals a distinct circular FS which arises from states within 10 meV of \EF\ observed  in figure \ref{gss}(f) at low temperature (T=7K) and weaker intensity wings extending to $\pm$0.5 \invA.
This corresponds to the high point of the $M$-band dispersion previously observed (and better resolved) at 31 eV \cite{Chatterjee2013,Boariu2013,Bareille2014} and 34 eV \cite{Yoshida2013}. 
In contrast at 42 eV, the angular map in figure \ref{gss}(e) exhibits  \EF\ intensity outside the $\pm$0.15 \invA\ hole band radius and is revealed in figure \ref{gss}(f) to arise also from a very flat band at \EF\ that extends out to $\pm$0.4 \invA. 

Above \THO, the \EF\ states at both 35 eV and 42 eV disappear as shown for the 40 K spectra and in the line spectrum comparisons in figure \ref{gss}(g). 
Temperature dependence at both photon energies was performed by cooling from $\sim$50 K down to 6 K (and for 35 eV also rewarming back to 50 K).  
The resulting \EF\ intensity line profiles for both 35 eV and 42 eV in figure \ref{gss}(h) show a rapid emergence of the \EF\ intensity \textit{above} \THO\ and only a weak rise or evolution below \THO.  
This behavior is generally consistent with the previous literature result of the $M$-band existing only in the HO state. 
However, in contrast to a reported evolution to zero \EF\ peak amplitude by 18K \cite{Yoshida2010,Yoshida2013}, our line spectra in figure \ref{gss}(g) still exhibit a weak \EF\ peak above background at $\sim$23K.  Our result is observed both for cooling and warming, as shown in the \EF\ intensity image in figure \ref{gss}(i), indicating that directional hysteresis of the $T$-reading is not an explanation for the discrepancy.  
This behavior is discussed later in section \ref{fsgap} in the context of a fluctuation regime. 
 Also from our photon-dependent mapping of \kz-dispersing states in figure \ref{gss}(d) indicating a \G-centered electron-like origin of these narrow states at \EF, we cannot claim that we are observing the appearance of zone-folded states. 
Zone-folding is further discussed in section \ref{zonefold}.

\subsection{X-point} \label{xpt}

\sectionmark{FIG. 10. X SS }
\begin{figure}[t]
\begin{framed}
\begin{center}
\includegraphics[width=15cm]{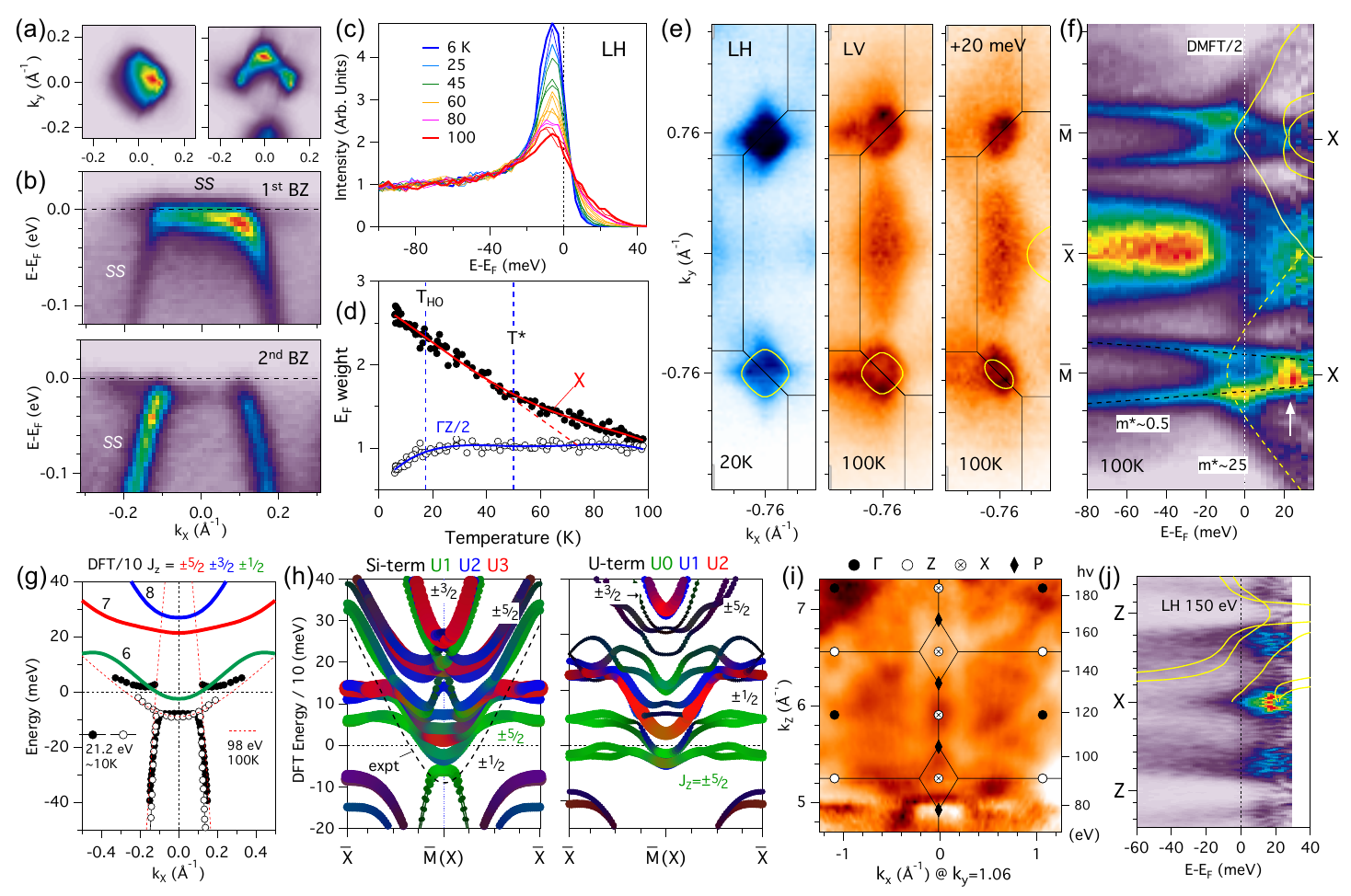}
\caption{
\textbf{X-point surface and bulk states.} 
On-resonance 98 eV (a) FS maps and (b) VB maps for first and second BZ X-points of the Si-terminated surface. 
(c) Temperature dependent spectra from the center of the X-point for LH polarization.
(d) $T$-dependent summary of the X-point  $f$-peak amplitude in relation to \Tcoh\ coherence and \THO\ hidden order temperatures, with comparison to the central \G Z/2 region.
(e) Fine-step constant energy maps at 20 K and 100K using LH and LV polarization showing surface-like four-fold symmetry at the X-point at \EF\, and a bulk-like two-fold symmetry for an energy cut at 20 meV above \EF.
Comparison is made to DFT X-point FS contours from the \TRS\ surface slab calculation and from a bulk \URS\ calculation.
(f) High temperature VB map divided by the Fermi-Dirac function for T=100K with fits to light effective mass (m*$\sim$0.5) hole and heavy (m*$\sim$25) electron bands dispersing above \EF. 
Arrow at +20 meV indicates the energy cut for the bulk-like two-fold symmetry map in (e).
Also overplotted are bulk DMFT bands rescaled by 0.5.
(g) Summary of He I literature X-point band dispersions, dots \cite{Boariu2013} and circles \cite{Zhang2018}, compared to the 98 eV 100K electron band fit and to bulk DFT bands rescaled by 0.1 and color-coded by J=5/2 J$_z$-character.
(h) DFT slab calculations of the X-point (\Mbar) region for three unit-cell Si- and two-unit-cell U-terminated \URS\ showing the relative energy shifts of the surface (U0), near-surface (U1), and bulk (U2,U3) electron-like $f$-states. 
(i) Photon-dependent \kxkz FS map showing bulk-like \EF\ intensity variations along the central -X-P-X-P-X- line.
(j) Z-X-Z VB map for a U-terminated surface at 150 eV (LH polarization) including division by the RC-FDD to highlight that the X-point gets relatively stronger intensity above \EF.
}
\label{xss}
\end{center}
\end{framed}
\end{figure}
 
Next we summarize our knowledge of the bulk and surface X-point electronic structure. 
Bulk band theory predicts a shallow electron pocket at the X point that compensates for the charge imbalance of the FS electron-sheets at \G\ and hole-sheets at Z.
For the \SiT surface, observation of such a bulk state might be obscured by the strong U 5$f$ surface state interior to the Si 2$p_z$ hole pocket that was identified in section \ref{spatial}.  
For the \UT surface, or aged or surface modified \SiT surfaces, where the hole pocket surface state is suppressed, we still see a persistent \EF\ weight at X for 98 eV on-resonance excitation, suggestive of a shallow bulk band.
However, for the higher photon-energy SX-ARPES measurements, Fujimori \etal\ characterizes small residual  spectral weight at the zone corner as having surface-like 2D character, and concludes that no bulk X-point pocket exists \cite{Fujimori2021}.

In the above observations, no shallow band energy minimum or \kF-crossing is resolved or characterized.  
Looking to lower photon energy, the X-point has been investigated using He I 21.2 eV excitation by Boariu \etal\ \cite{Boariu2013} and by Zhang \etal\ \cite{Zhang2018} for the \SiT\ surface, identifiable by the presence of the hole band surface state. 
Boariu \etal\ observe, from 10K to 34K, the hole band to terminate below \EF\ with flat spectral weight both interior to the hole band at -8 meV and just above \EF\ outside the hole band. 
They model this as a shallow electron band with $\sim$2 meV depth that hybridizes with the hole band creating an $\sim$11 meV gap that persists all the way up to 35K.  
While both bands were assumed to have bulk origin, such strong hybridization with a surface state hole band strongly suggests a surface origin also of the unhybridized shallow electron pocket.
In contrast, Zhang \etal\ observes the surface state hole band to terminate (with no gap) at a shallow electron band with -9 meV band minimum and \kF$\approx\pm$0.24 \invA\ outside the $\pm$0.1 \invA\ hole band width. They infer hybridization between the two states, but no gap is observed.
The gapped X-point spectra for the Si-terminated surfaces is also observed at 50 eV by Bareille \etal\ \cite{Bareille2014} where PM and HO phase FS maps, measured at 20K and 1K, show a distinct change in symmetry.  While a $T$-dependent change suggests a bulk origin, the observed two symmetries do not agree with the bulk DFT prediction.

We further explore in figure \ref{xss} the X-point electronic structure of the \SiT surface returning to the \dtof\ resonance conditions and make comparison to DFT surface slab calculations of \URS.
First, the strong intensity flat band of $f$-weight confined to the \textit{interior} of the X-point surface-state hole band, observed in section \ref{spatial}, is reproduced in figure \ref{xss}(a) and (b) for the \SiT surface using the on-resonance 98 eV excitation with p-polarization (LH) of the incident x-rays.
In addition, a dramatic  suppression of this $f$-state intensity is observed 
by measuring another X-point  in the second BZ at larger angle. 
With the flat interior $f$-state intensity suppressed, another weaker intensity narrow band is revealed with a shallow electron dispersion that extends \textit{outside} the X-point surface state hole band, similar to the Zhang \etal\ He I ARPES.
This suggests either a single band with a strong polarization difference interior versus exterior to the hole band, or     the existence of \textit{two} separate narrow bands that overlap at the X-point.

Another method of suppression of the X-point surface-state $f$-peak, in order to further investigate the shallow electron band on the \SiT surface, is to exploit its strong temperature dependence, 
previously reported in early studies by Denlinger \etal\ \cite{Denlinger2002} where the possibility of a surface state origin was not discussed.
In figure \ref{xss}(d), a fine step temperature dependence from 6 K to 100 K of the X-point spectra for 98 eV with LH polarization confirms such a dramatic $T$-dependence amplitude suppression starting from a large  $>$4:1 peak-to-background ratio at low $T$. 
A quantitative analysis of the strong \EF-weight variation (with background subtraction and normalization to 1 at high $T$) versus temperature is plotted in figure \ref{xss}(d). 
Upon cooling from high $T$ the approximately linearly increasing $f$-weight is observed to be insensitive 
to the bulk hidden order transition at \THO=17.5 K,  consistent with it being a non-bulk surface state. 
For comparative contrast,  a reduction of \EF-weight beginning at or slightly above the bulk HO transition is observed at the intermediate \G Z/2-point between the two X-points.

Next, low and high temperature fine-step FS maps containing two X-points, measured on-resonance at 98 eV,
are presented in figure \ref{xss}(e).  
At 20K, the four-fold symmetric diamond shape of strong \EF\  $f$-state intensity contrasts with the two-fold symmetry of the bulk BZ at the X-point and is consistent with the size of the \TRS\ DFT surface slab X-point FS contour.
At 100K,  the weakened $f$-state intensity shows up as a depletion of the interior of the 4-fold symmetric diamond  from the Si-terminated surface state hole band. Also the narrow band intensity exterior to the diamond appears extended from the corners of the diamond shape along $<$100$>$ directions, in agreement with the Bareille \etal\ observation \cite{Bareille2014}, and  inconsistent with a bulk DFT prediction of an electron pocket elongated along the bct BZ boundary (i.e. along Z-X-Z).
Interestingly, though, such a bulk-like two-fold symmetric X-point intensity \textit{does} appear at a constant energy cut 20 meV above \EF, also shown in figure \ref{xss}(e).

To investigate this result further, a VB map view of these high-$T$ X-point dispersions is shown in figure \ref{xss}(f), with visualization of the states up to 40 meV ($\sim$5\kB $T$) above \EF\ enhanced by dividing the spectra by a resolution convolved Fermi-Dirac distribution function (RC-FDD).
The shallow electron pocket dispersion can be now be traced up to 30 meV above \EF, and can be fit to a non-parabolic dispersion with an effective mass of m*$\sim$25m$_e$.   
The light hole band dispersion is also observed to extend above \EF\ with a parabolic effective mass of m*$\sim$0.5m$_e$ below \EF, but is then observed to close with a shallow maximum at only +20 meV, which is the source of the two-fold symmetric intensity in the +20 meV intensity map. 
The shallow electron pocket dispersion fit with a -9 meV band minimum, is then in figure \ref{xss}(g), compared to and found to be consistent with the band energies derived from the two He I measurements.   

A possible interpretation of this experimental result is that the true bulk electron band minimum is 20 meV above \EF, and $both$ the high intensity $f$-state interior to the surface state diamond and the narrow $f$-weight extending outside the diamond are of surface origin.
However, the bulk DFT+DMFT  X-point band structure, overplotted in figure \ref{xss}(f) with an additional 2$\times$ energy renormalization, and the bulk DFT  bands,  plotted in figure \ref{xss}(g) with 10$\times$ renormalization (DFT/10), show the existence of the next highest energy bulk bands also at approximately +20 meV. 
If the experimental +20 meV state corresponds to such bulk bands with DFT $f$-orbital character of $j_z$=$\pm\frac{3}{2}$ or $\pm\frac{5}{2}$, then there should be a bulk band with $j_z$=$\pm\frac{1}{2}$ character in the vicinity of \EF, albeit with a shallower -2.5 meV prediction compared to the observed -9 meV minimum.

Next, we look at the predictions of DFT/10 surface slab calculations for both surface terminations of \URS\ including the U 5$f$ states. For the \SiT surface in figure in \ref{xss}(h), a 3 unit cell slab containing three U sites is used to better assess the bulk-like energy convergence between the two innermost U sites. 
The slab calculation very nicely predicts the flat $f$-weight originating from the near-surface U1 atomic site ($\pm\frac{1}{2}$ character) confined to the interior of the surface Si  $p_z$  hole band. 
Also bulk-like U2 and U3 site electron bands, with nearly identical energies, exist at approximately +20 meV above \EF\ with $j_z$=$\pm\frac{3}{2}$ and $\pm\frac{5}{2}$ character, similar to the bulk DFT/10 calculation.
The electron band crossing below \EF\  is again shallower in energy than the experiment (dashed line curve) and, interestingly, has \textit{mixed} U1 ($j_z$=$\pm\frac{5}{2}$)  and U2 ($j_z$=$\pm\frac{1}{2}$) character.
The similar experimental effective mass curvature of the electron band crossing below \EF\ confirms the suitability of the DFT/10 energy scaling.
Also, the most interior U3 ($j_z$=$\pm\frac{1}{2}$) band minimum is different from U2 and slightly above \EF. 
A systematic rigid Fermi-energy shift to align the U2 band to the experimental energy would bring the bulk U3 band below \EF, close to the bulk DFT prediction.   
Hence, we conclude from the slab theory that the experimental electron band crossing below \EF\ at X likely originates from a rather deep near-surface region involving $two$ sub-surface U layer 5$f$-states, and that the probe-depth is insufficient to obtain spectral intensity from the more bulk-like U3 site  that is  $\sim$12 \AA\ (1.25$c$) below the surface.    

Hence we turn our final X-point focus to the comparison of theory and experiment for the \UT surface, where the complexity of the  Si $p_z$ dangling bond hole band is removed.  
A two unit-cell surface slab calculation in figure \ref{xss}(h), predicts both U2 and U1 interior sites to have similar converged electron band energies both at +20 meV and slightly above \EF\ at X. 
In this case, the less-coordinated U0 surface atoms exhibit a narrower bandwidth and split-off an electron-pocket just below the bulk-like electron bands. 
Disruption of the long range order of the U0 site by disorder, vacancies, \SiT terraces, or surface adsorption can diminish the intensity and clarity of this surface state electron pocket, and result in the broad $k$-independent $f$-spectral weight near \EF\ that is experimentally observed. 
The persistent X-point $f$-spectral weight at 98 eV, without the presence of the hole-band surface state, is observed at higher photon energies to be periodic at the bulk X-points along $k_z$, as highlighted in the figure \ref{xss}(i) focus on the vertical X-P-X-P-X line.
The 150 eV Z-X-Z cut from this data set, also plotted figure \ref{vb150}(f), is shown in figure \ref{xss}(j) with the division by the RC-FDD to enhance the states above \EF. What is observed is that the X-point peak becomes relatively stronger above \EF\ compared to the Z-points. 
While this is suggestive of the bulk X-point minimum just touching or being above \EF, high-resolution ARPES of the X-point for the \UT surface, 
at elevated temperatures and without weak spectral contamination from \SiT regions, e.g. as present in figure \ref{xy}(c),
would be required to better quantify the true bulk X-point band minimum.  

Quantum oscillations (QO) also have the potential to identify a bulk X-point FS orbit. 
Being restricted to very low temperatures in the HO and AFM phases of \URS, due to heavy mass amplitude damping \cite{Ohkuni1999}, QO have nicely observed FS orbit angular dependences consistent with the zone-folded breakup of the PM phase \G\ and Z large FS sheets into smaller sheets \cite{Hassinger2010}.
To date, no small size QO orbit has been assigned to the zone-folded X-point (tetragonal M-point). 
However, the complementary cyclotron resonance technique,  which can directly measure the hierarchy of relative effective masses of FS orbits  and their angular dependences, albeit without quantification of the orbit size, \textit{has} made an assignment to the tetragonal phase M-point \cite{Tonegawa2012}. 
After making cyclotron mass assignments to DFT FS sheets similar to the QO $\alpha, \beta$ and $\gamma$ FS orbits, an additional heaviest cyclotron mass state is observed to have a splitting along the (110)  direction that is consistent with expectations for the DFT zone-folded X-point.

\subsection{N-point region } \label{npt}  
  
\begin{figure}[t]
\begin{framed}
\begin{center}
\includegraphics[width=15cm]{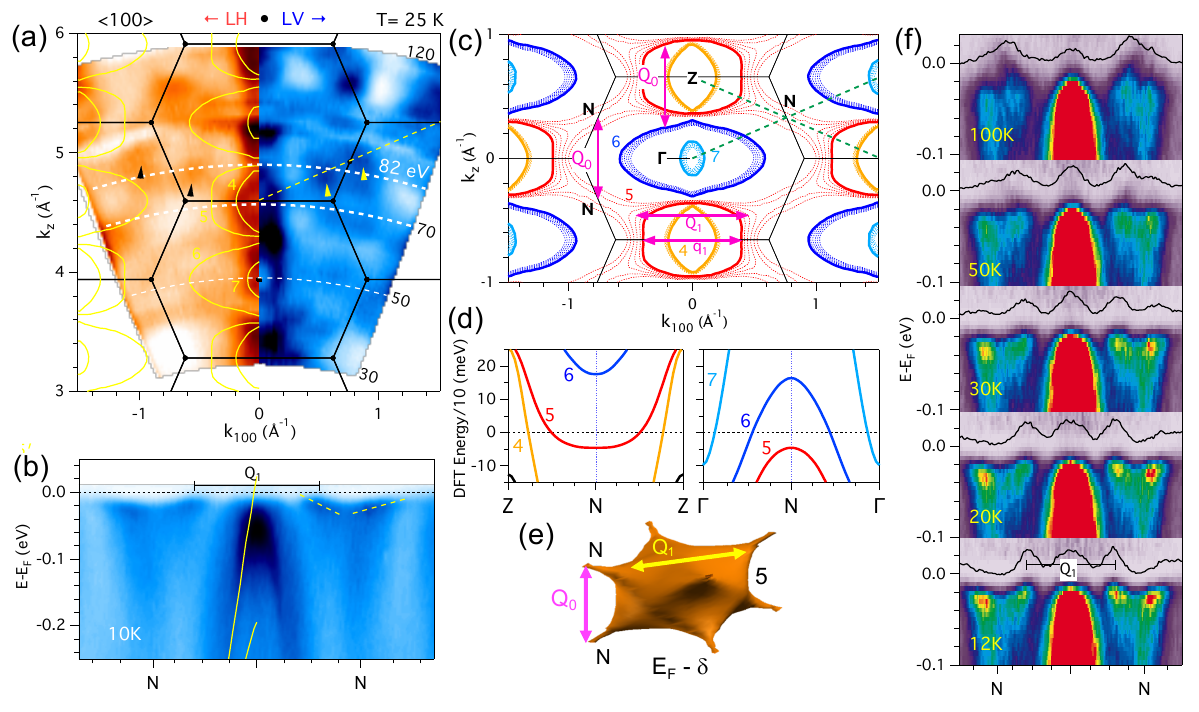}
\caption{
\textbf{N-point region electronic structure.}
(a) Normal emission \kxkz\ photon dependent FS maps in the \adir-\cdir\  plane with both LH (left) and LV (right) x-ray polarization with overplotted DFT FS contours (left).  
A constant photon energy arc is overplotted for 82 eV that cuts through the two maximum intensity features (arrows) symmetric about the N-point along Z-N-Z.
(b) Low temperature (10K) 82 eV VB map measured with LV polarization and with comparison  to the Q$_1$ scattering vector.
(c) DFT \kxkz\ FS contours for the $<$100$>$ direction including 10 meV step constant energy contours down to \EF-50 meV, illustrating how a long Z-N-Z neck develops for band 5 at only -40 meV.  
Various candidate Q$_0$ and Q$_1$ scattering vectors are also shown (see text).
(d) DFT Z-N-Z and \G-N-\G\ bands (renormalized by 10$\times$) illustrating the saddle point nature of the shallow band 5 neck.
(e) Schematic of the -40 meV isoenergy surface of DFT band 5 showing the elongated N-point necks with inherent Q$_0$ vector separation. 
(f) Series of 82 eV VB maps (and \EF\ momentum profiles) at selected temperatures upon cooling the sample from 100K to 12K.
}
\label{hv82}
\end{center}
\end{framed}
\end{figure}

Finally, we widen our global photon-dependent FS map in the  \adir-\cdir\ plane beyond the narrow normal emission region discussed in section \ref{gzpt}.  Figure \ref{hv82}(a)  shows \kxkz FS maps for \SiT \URS\ for 30-150 eV for both LH and LV polarizations. 
This plane does not cut through the X-points, so those surface states are avoided. 
Away from the central normal emission region, our attention is drawn to the other \EF\ intensity maxima near the staggered BZ boundary, 
and in particular to the two states on either side of the N-point (marked by arrows in figure \ref{hv82}(a)). 
These states are visible for both \SiT and \UT cleave surface regions and also for both LH and LV polarizations,  with the sharpest most distinct spectra coming from the \SiT surface with LV polarization.

\sectionmark{FIG. 11. N-point }

The two zone-boundary states  appear symmetrically about the N-point in adjacent BZs along the the diagonal Z-N-Z line, and are best (but not ideally) intersected by the 82 eV photon energy arc.
The 82 eV band dispersion cut, measured in the HO state using LV polarization in figure \ref{hv82}(b), reveals the origin of the \EF\ intensity maxima to be a pair of states with heavy mass band velocities of \vF\ $<$ 0.1 eV-\AA.
The low $T$ sharp QP peak maxima are also noted to have a momentum separation of 0.6a* which matches the incommensurate wavevector observed by inelastic neutron scattering.
The long vertical waterfall intensities to higher binding energy are interpreted as coming from electron correlated self energies 
or inelastic loss effects, 
and not as light effective mass bands which do not exist at the N-point in the \TRS\ band structure.

The location of these N-point states appears to be well outside the predicted DFT FS contours overplotted onto the  \kxkz FS maps. 
Such disagreements with the DFT FS observed in sections \ref{urs200} and \ref{sxarpes} were rationalized before as being the effects of inadequate experimental energy resolution combined with the lack of theoretical energy renormalization of the extremely heavy mass regions of $k$-space.   
While the improved energy resolution here of $<$15 meV rules out that source of the discrepancy, we are nonetheless similarly motivated to explore the DFT shallow BE contours just below \EF. 
In figure \ref{hv82}(c) the DFT FS contours for this \kxkz cut are plotted with band index color-coding and with dashed line BE contours down to \EF-50 meV in 10 meV steps of the DFT energy scale. 
(Note that this is 4$\times$ smaller than the 0.2 eV depth of the BE contours plotted in figure \ref{urs150}(e).)  
While the FS contour shapes are preserved and sizes are not much changed for bands 4, 6 and 7 sheets over this small energy scale,  the contours of the hole band 5  are observed to have a dramatic change in shape and size due to its formation of a very  shallow energy neck connecting the Z points through the N point. 

The DFT band 5 dispersion along Z-N-Z, with a 10$\times$ renormalized energy scaling applied in figure \ref{hv82}(d), exhibits a very shallow dispersion confined to less than 5 meV bandwidth over a very large $\pm$1 \invA\ range along the neck, while the \G-N-\G\ dispersion cut shows a hole like dispersion at N, demonstrating a saddle-point nature of this band 5 neck.   
Hence, we naturally conclude that the experimental heavy mass zone boundary states in figure \ref{hv82}(a) and (b) must be coming from band 5. 

 An incommensurate Q$_1$ scattering vector, noted in the experimental 82 eV VB and FS maps,
is also overplotted onto the DFT energy contours in figure  \ref{hv82}(c).  
While the straight vertical edges of the DFT band 5 FS contour would be attractive for a FS nesting scenario, the Q$_1$ vector length of 0.91 \invA\ is observed to be slightly larger and inconsistent with the FS contour  separation of q$_1$=0.8 \invA. 
Plotted at the approximate location of the 82 eV measurements, the Q$_1$ vector appears to correlate well to the tips of this vertical segment of the band 5 FS where the shallow necks emerge towards the N-point. 

Two commensurate Q$_0$=(0,0,1) scattering vectors are also overplotted onto the DFT energy contours in figure  \ref{hv82}(c).
The first one corresponds to a FS nesting scenario previously proposed in the literature \cite{Oppeneer2010,Oppeneer2011,Ikeda2012} that links the upper (or lower) edges of the band 6 \G\ electron FS and the band 5 Z-point hole FS where the curvatures have a very good match over an extended region of $k$-space. 
The BE contour analysis above immediately suggests an additional point-like contribution to Q$_0$=(0,0,1)  nesting coming from scattering between N-points which inherently nests $all$ energy bands due to the crystal symmetry.

These two newly proposed Q$_0$ and Q$_1$ $intra$-band scattering vectors are graphically illustrated in relation to the band 5 three-dimensional DFT isoenergy surface 4 meV (DFT/10) below \EF\ in figure \ref{hv82}(e).
This inherent nesting property combined with the presence of narrow $f$-states at a small few meV energy scale below \EF, 
which are thermally active in the $T$ range of the HO transition, suggests a possible relevance to the HO physics.

The basic experimental $T$-dependence of these heavy-mass N-point bands is presented in figure \ref{hv82}(f) for 82 eV VB maps at five select temperatures between 12K and 100K. The spectra are divided by a RC-FDD function and employ a common color table that saturates the central states to better highlight the thermally excited states above \EF\ at higher $T$.
What is observed is that the spectra for the three lowest temperatures of 30K, 20K and 12K are essentially the same, except for some sharpening of the Q$_1$-separated heavy mass QP peak.  
Only above 30K do we see the distinct $T$-evolution of a curious \textit{reverse} hole-like dispersion above \EF\ centered on the N-point.
This exotic $T$-dependence is not straightforwardly mapped onto textbook two-band $f$-$c$ Kondo hybridization scenarios, especially since there are no theoretical occupied non-$f$ conduction bands in this region.
The apparent saturation by 30K of the N-point $T$-evolution upon cooling, i.e. above \THO, is notably consistent with the 42 eV and 35 eV $T$-dependent behavior above \THO\ found in section \ref{gzpt}.    
A detailed analysis and interpretation of this N-point $T$-evolution is presented elsewhere \cite{Denlinger2022}.

\section{Discussion} \label{discussion}

Here we provide some additional commentary on the results presented above and on specific issues discussed in the literature.

\subsection{Ordered phase zone folding} \label{zonefold}

Symmetry breaking of the electronic states in \URS\  from D$_{4h}$ symmetry in the PM phase to C$_4$ symmetry occurs for both the HO and AF phases. 
The associated transformation of the body-centered tetragonal to simple tetragonal Brillouin zone, as shown in figure \ref{bz}(b), results in a zone folding of the electronic states.  
Direct evidence of this symmetry reduction is given by Raman spectroscopy with polarized light 
identifying a 1.8 meV excitation in the HO phase with A$_{2g}$ symmetry that is not allowed for D$_{4h}$ symmetry  \cite{Kung2015,Buhot2014}. 
Also the similarity of quantum oscillation FS orbit angular dependences between HO and AF phases  provides additional evidence for the zone-folding in the HO phase \cite{Hassinger2010}. 
A primary focus of the high resolution ARPES studies has been on a particular HO phase $M$-shaped band whose presence is attributed the zone-folding effect.
The 7 eV laser ARPES study \cite{Yoshida2013} reported that the $M$-shaped band at the zone center only exists in the HO phase and disappears by 20K. 
Subsequent variable photon energy studies measured the $M$-shaped band feature at 31 eV \cite{Bareille2014} or 34 eV  \cite{Yoshida2013}, ascribed to the Z-point location, and compared the results to \G-point measurements at either 17, 19 or 50 eV, and to measurements above \THO. 
Both groups conclude that the $M$-shaped band, observed in the HO phase at both Z and \G,  is a transformation of a shallow electron band at the Z-point above \THO\ with nothing specific observed at \G\ in the PM phase.

\begin{figure}[t]
\begin{framed}
\begin{center}
\includegraphics[width=15cm]{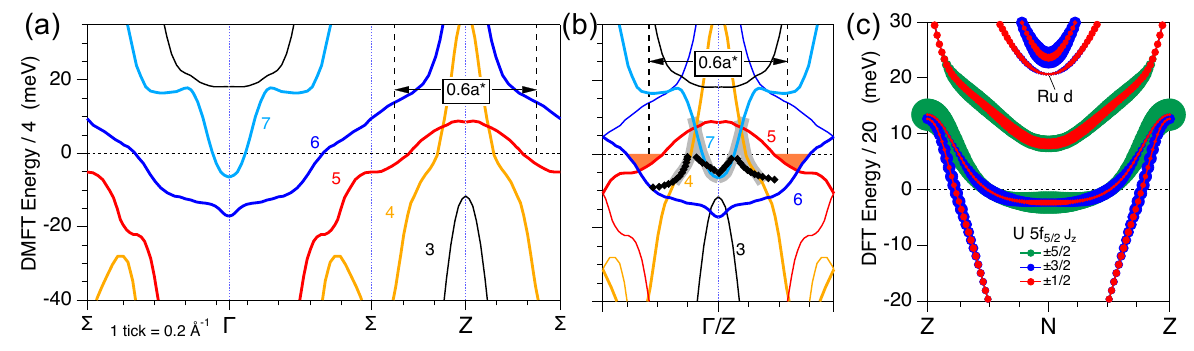}
\caption{
\textbf{Ordered phase zone-folding}  
(a) DMFT PM phase calculated bands along (100) \G-Z with band color-coding consistent with previous plots.
(b) Zone-folding of the DMFT PM bands simulating the ordered tetragonal phase, and illustrating the band 4 and 7 overlap origin of the $M$-shaped band dispersion, and band 5 and 6 overlap origin of shallow pockets at larger momenta.
The relationship of the Q$_1$=0.6a* incommensurate vector length to the BZ ($\Gamma$-$\Sigma$) and to PM and zone-folded features is illustrated in (a,b).
The DMFT energy scale has been further renormalized by 4$\times$ to provide agreement with the experimental $M$-band dispersions (dots) from \cite{Chatterjee2013} (left) and \cite{Bareille2014} (right). 
(c) Z-N-Z DFT band structure (with 20$\times$ energy renormalization) showing the large-momentum flat band extent and $J_z$ $f$-orbital character of the N-point region.
}
\label{zfold}
\end{center}
\end{framed}
\end{figure}

It is instructive to compare this interpretation to the theoretical band structure predictions.
Figure \ref{zfold}(a) and (b) show the DFT+DMFT band structure along \G-Z in the PM phase and those same bands zone folded such that \G\ and Z are equivalent.
It is observed that an $M$-shaped band naturally occurs from the overlap of the band 4 small Z-point hole dispersion, forming the outer edges of the $M$, with the band 7 small \G-point electron pocket forming the inner part of the $M$. 
While the momenta of the theoretical $M$-band are in decent agreement with the ARPES measurements, overplotted in figure \ref{zfold}(b),   obtaining occupied bandwidth agreement requires an additional 4$\times$ reduction of the DMFT energy scale, i.e. a net 20$\times$ energy renormalization compared to the DFT calculation.   

It is noteworthy that the \G- and Z-origins of the zone-folded components of the $M$-band are opposite between theory and the ARPES results.  One possible explanation of this discrepancy lies in possible low-photon-energy final-state effects in the photoemission process whereby dipole transitions from the Fermi level to the unoccupied band structure are not allowed, due to selection rules or gaps in the final-state bands, which can result in ghost bands, \kz-shifted bands and other artifacts \cite{Strocov2003}. These effects become less likely at higher photon energies where the unoccupied final states occur more uniformly in the reduced zone scheme.  
For example, in figure \ref{gss}(c) and (d) it is observed that high intensity points in the photon dependent maps for LV polarization agree well at 80 eV and 60 eV with the DFT hole-to-electron crossover points between bands 4/5 and bands 6/7, whereas the strong enhancement at 35 eV is shifted from the expected energy of about 40 eV.

\sectionmark{FIG. 12. FS nesting and Zone-folding }

\subsection{Origins of heavy effective mass} \label{heavymass}

DFT theory is well known to underestimate electron correlations and predict 4$f$-band widths orders-of-magnitude too large in rare-earth compounds.  Examples of large 4$f$ renormalization factors between DFT and DFT+DMFT calculations include 100$\times$ in CeIrIn$_5$ \cite{Choi2012} and 10$\times$ in SmB$_6$ \cite{Kim2014}, with some validating correspondence from experiment.
The magnitude of the $f$-renormalization factor and the degree of proximity of the states to \EF,  are key parameters in the itinerant versus localized characterization of the system.
The $f$-renormalization factor reflects the strength of the hybridization to non-$f$ bands or the possibility of $f$-``bands'' from direct $f$-$f$ hopping (which is unlikely for ternary compounds with large U-U distances). 
Due to the larger atomic radial spread of 5$f$ orbitals compared to 4$f$ orbitals, a much smaller renormalization factor is expected for \URS\ and even a factor of unity has been proposed \cite{Oppeneer2010}. 
Thus, the renormalization factor of 5$\times$ for the DMFT calculation presented here seems reasonable, so it is quite surprising that the experimental value from figure \ref{zfold}(b) of 20$\times$ is even greater than that of SmB$_6$. 
The DFT renormalization factor also may not be uniform throughout the BZ, as evidenced by the 10$\times$ factor used in the X-point analysis in section \ref{xpt} which provided good agreement to both the shallow electron band dispersion and a higher energy bulk state 20 meV above \EF.

Also, for large renormalization factors, agreement of DFT FS details with experiment is not a general rule, despite the Luttinger electron counting theorem, as exemplified by the case of CeCoIn$_5$ \cite{Jang2020}, where the too large DFT $f$-bandwidth is too disruptive to sensitive non-$f$ bands having a low-energy scale relation to \EF\ prior to Kondo hybridization.  
However in the case of \URS, the non-$f$-band crossings of the $f^0$-equivalent compound \TRS\ have rather light effective masses and wide separation between the Z-point and its zone boundary, hence are less sensitive to such DFT U 5$f$-band disruption.
Thus good agreement was found in figure \ref{vb150}(c) between the DMFT \kF\ values and the 5$\times$ scaled DFT bands.
It is also noteworthy that for fixed $f$-$c$ hybridization strength, the key recipe here for heavy effective band masses \textit{is} the wide separation between the small and large \kF\ hole bands of \TRS.  For example, the shallow $f$-band velocities of $<$0.1 eV-\AA\ along Z-X, that show up as lines of high $f$-weight in the SX-ARPES maps, represent a dispersion from slightly above to slightly below \EF\ over a span of 0.6 \invA.
Even more extreme is the span of nearly 1 \invA\ along the Z-N-Z neck in figure \ref{hv82}(d) where the occupied shallow $f$-band remains $<$5 meV below \EF.

\subsection{Itinerant versus Localized $f$-states} \label{CEF}

A common binary sorting criterion to distinguish between models of the hidden order in \URS\ is that of localized $f^2$ origins versus the itinerant $f^3$ origins.  
Also experimentally, a localized-itinerant duality is observed between the probes of local order, e.g. non-resonant inelastic x-ray scattering (NIXS), that tend to reveal signatures of $f^2$ character versus the longer range probes, e.g. hard x-ray photoemission (HXPES) and ARPES, which reveal itinerant character  \cite{Amorese2020}.  A conceptual dilemma exists in how much itinerant $f$-$c$ hybridization can exist before the $f^2$ CEF physics gets washed out, with the $f$-occupation of n$_f\sim$2.5 being a rough threshold value. 
Experimentally, inelastic neutron spectroscopy fails to observe a distinct $k$-independent CEF excitation peak in \URS.
A dual character scenario that gets around this dilemma is also proposed in which two $f$-electrons are localized and exhibit the $f^2$ CEF physics while the remaining $\sim$0.7 electrons participate in the itinerant Kondo hybridization \cite{Zwicknagl2003}.  

The zone-boundary X-point and N-point regions highlighted in section \ref{tdep} offer an interesting viewpoint on these itinerant versus localized versus dual character scenarios.  
Due to the lack of interference of hybridizing non-$f$ bands over large momentum spans,
they represent fairly clean views of the DFT-predicted CEF energy splittings of the $f$-states. 
At the X-point a $J$=$\frac{5}{2}$ three-band hierarchy of fairly pure $J_z$=$\pm\frac{1}{2}$,  $\pm\frac{5}{2}$ and $\pm\frac{3}{2}$ $f$-orbital character states, with $\sim$20 meV first-excited-state separation for 10$\times$ DFT energy scaling, was shown in figure \ref{xss}(g). 
In contrast, the N-point $f$-states exhibit much stronger binary mixing of $J_z$ orbital characters in figure \ref{zfold}(c), including a distinct $k$-dependent variation of the mixing of the $J_z$=$\pm\frac{5}{2}$ and $\pm\frac{3}{2}$ states along the band below \EF.
Plotted with a 20$\times$ renormalization of the DFT energy scale in figure \ref{zfold}(c), the next highest $f$-orbital is still $>$10 meV ($>$120K) above the shallow occupied N-point $f$-band. 
This DFT prediction is inconsistent with a scenario of two narrowly split f$^2$ singlet CEF levels with thermally-activated mixing relevant to the HO transition temperature  \cite{Kung2015,Kung2016}. 
Instead, it is the thermal excitations to \EF\ of this \textit{single} very shallow binding energy N-point $f$-band that are relevant to the HO temperature scale.

\subsection{Commensurate and Incommensurate Scattering} 

Inelastic neutron scattering observes strongly dispersive spin fluctuation excitations with scattering vectors of Q$_0$=(0,0,1) and  Q$_1$=(1$\pm$0.4,0,0) and with energy gaps of 1.8 meV and 4.5 meV, respectively \cite{Broholm1987,Wiebe2007}. 
From the DFT band structure, a Q$_0$ FS nesting condition between the matching curvature of the large \G-centered  electron FS and the large Z-point hole FS  has been noted \cite{Oppeneer2010,Oppeneer2011,Ikeda2012}. 
In addition, various other FS nesting scenarios for the incommensurate Q$_1$ vector have been proposed.
Elgazzar \etal\ \cite{Elgazzar2009} noted a 0.4a* separation between Z-point hole \EF\ contours of bands 4 and 5, and thus proposed a 0.6a* nesting between band 5 and the zone-folded band 4 in the next BZ.  
Kawasaki \etal\ \cite{Kawasaki2011} proposed a direct 0.6a* nesting between the hole bands 4 and 5 in the same BZ based on their SX-ARPES derived star-shaped band 4 and larger circular band 5 FS contours. 
Bareille \etal\ \cite{Bareille2014} noted that the Q$_1$ vector spans the distance between the centers of the electron pocket `petals' formed from the zone-folding of bands 5 and 6, as illustrated in figure \ref{zfold}(b).

Here we have presented in section \ref{npt} experimental and theoretical evidence for alternative Q$_0$ and Q$_1$ scattering vectors deriving from the N-point electronic structure.
A key implication of the existence of two different Q$_0$ interband and intraband nesting conditions 
is the possibility that they provide the competition that results in the two different HO and AF ordered groundstates of \URS. 
Also, the experimental Q$_1$ scattering $k$-points at $\pm$0.3a* are observed to result from a $T$-evolution that is ``arrested'' just above \THO\ thereby preventing the DFT groundstate FS from being achieved where the band 5 N-point necks emerge.

\subsection{Fermi surface gapping} \label{fsgap}

From a thermal activation fit of the specific heat profile below \THO\ and from the ratio of the specific heat below and above \THO, it was estimated that an HO energy gap of $\sim$11 meV develops over 40\% of the Fermi surface \cite{Maple1986}. 
Far-infrared spectroscopy shows a low energy 6 meV gap develop below \THO\ \cite{Bonn1988} as well as spectral weight transfer on the 16 meV scale associated with hybridization gapping and the onset of coherence around 75K \cite{Bachar2016}.
Point contact tunneling spectroscopy observes an 11-14 meV gap opening with a mean-field onset above \THO\ at $\sim$22-26K \cite{Samuely1995,Rodrigo1997,Lu2012,Park2012}, but also as high as 34K \cite{Park2014}, that is interpreted \cite{Park2012,Park2014,Zhang2020} as hybridization gapping associated with Kondo coherence, and thus implying an insensitivity of the technique to any signature of a HO/AF related gap.  
Scanning tunneling spectroscopy observes both the slow development of a broad asymmetric Fano lineshape profile 
over a 100 meV energy scale beginning at $<$100K that is associated with Kondo coherence, and a much narrower 4-5 meV gap at \EF\ with a BCS-like $T$-dependence onsetting around 16K \cite{Aynajian2010}.  

For $k$-resolved FS gapping, scanning tunneling quasiparticle interference (QPI) is able to visualize a $q$-resolved gapping interaction between light and heavy mass band(s) at a scattering wavevector of $q$ = 0.3a* (2$\times$ smaller than the $Q_1$ vector) along both (100) and (110) directions, but with a complexity that requires modeling by interband scattering between the two different DFT hole bands \cite{Yuan2012}. 
DFT band theory employing the AF tetragonal BZ and hybridization of zone-folded bands makes a specific prediction of the breakup of the large PM phase FS sheets centered on the Z- and \G-points into four smaller disconnected FS `petals' along (100) whose predicted quantum oscillation angular dependence agrees well with experiment  \cite{Hassinger2010}. 
Furthermore, one high resolution ARPES study in the \G-plane at 50 eV photon energy reports observation of this `petal' FS reconstruction resulting from experimental 5 meV gapping along (100) and 7 meV along (110) directions \cite{Bareille2014}.


In addition to the above zone-folding gapping effects, various experiments with fine-step $T$-variation give evidence of the onset of gapping behavior occuring just above the hidden order phase and the proposal of a ``pseudogap'' regime with a 25-30K crossover temperature \cite{Haraldsen2011,Shirer2013} that is distinct from the generic Kondo lattice hybridization gap with a higher 60-75K coherence $T$ onset, and that may or may not be a precursor to the HO transition.   
Fine $T$-step far-infrared spectroscopy exhibiting a $\sim$30K onset temperature for a higher energy scale spectral weight gapping associated with $f$-$c$ hybridization \cite{Levallois2011} was initially part of this precursor regime discussion, but the experimental onset of this behavior was later revised to be $\sim$75K \cite{Bachar2016} consistent with the resistivity maximum signature of the onset of coherence.   
Similarly, the initial $\sim$22-26K onset of a tunneling gap in quasiparticle scattering point contact spectroscopy was part of the pseudogap regime discussion, but an increased onset value to $\sim$34K \cite{Park2014} (still well below $T$*) and independence of the doped  \URS\ groundstate \cite{Zhang2020} has led to a hybridization gap interpretation dissociated from any precursor link to the hidden order.   
 
Quantitative $T$-dependent FS gapping ARPES studies at specific $k$-points with fine enough $T$ steps to distinguish the existence (or not) of a pseudogap regime are challenging and have not been reported.   
Here we have reported fine $T$-step characterization of other spectral features, such as (i) the QP peak \EF\ amplitude in the normal emission region at 35 eV and 42 eV in figure \ref{gss}(h), (ii) \EF\ spectral weight reduction at the intermediate point between two X-points in figure \ref{xss}(d), and (iii) \kF-shifting below 100K and the development $<$30K of narrow band QP peaks along Z-N-Z with $Q_1$ incommensurate wavevector separation in figure \ref{hv82}(f). 
In all three cases, the existence of a transition region above \THO\ is suggested by the characteristic behavior starting above \THO, but below $\sim$30K.   
Thus we believe the ARPES data presented here provides additional evidence for a fluctuation regime just above \THO\  that precedes a sharp ordering transition.

\subsection{Chemical doping}  \label{cdope}

Tuning of pressure, magnetic field and chemical composition has been used to help understand the range of existence and hence origins of the HO phase of \URS\ by revealing phase diagrams that suppress (or enhance) the HO state and also promote ferromagnetic or antiferromagnetic ground states.    While pressure and applied magnetic fields are incompatible with ARPES measurements, chemical doping such as Ru substitution is accessible. 
There are well characterized phase diagrams for Rh \cite{Yokoyama2004}, Re \cite{Bauer2005} and Fe \cite{Das2015} substitution of Ru, as well as for P substitution on the Si site \cite{Gallagher2016}.
ARPES measurements of 3\% Rh-doped \URS, where HO is suppressed and AF order exists below 10K, have been performed at the  two photon energy extremes of laser-ARPES \cite{Yoshida2010,Yoshida2011} and SX-ARPES \cite{Kawasaki2011b}.  
While no notable changes in the PM electronic structure are detected in either study, the narrow $M$-shaped \Gbar\ state observed in the HO state and interpreted as a signature of zone-folding,  is surprisingly absent in the Rh-substituted AF state at 7K where unit cell doubling is also expected to occur.
In contrast, a very recent high-resolution ARPES study of 5\% and 10\% Fe-doped \URS\ \cite{Frantzeskakis2021}  
does observe the same 31 eV Z-point $M$-shaped band also in the AF state, as expected, but with only subtle differences, such as a smaller size electron-pocket inner part  of the $M$-shaped dispersion. 
Also new attention was paid to spectral differences for the two different surface terminations including Si 2$p$ core-level characterization \cite{Frantzeskakis2021}. 

The N-point region, with its extremely shallow binding energy $f$-band, should be especially sensitive to such chemical doping perturbations, and tuning of the thermal activation of this region or its relative Fermi-edge DOS has great potential as a unifying explanation of the various phase diagrams.   
Investigation of the doping dependence of the N-point ARPES electronic structure is presented elsewhere \cite{Denlinger2022}.

\section{Summary}

In summary, we have given an overview of past ARPES results for the photon energy regimes of Fig. 1(c) and presented new results for the less studied intermediate low photon energy regime of 30-150 eV.  
The extreme surface sensitivity of this regime 
is the key tradeoff for the benefit of better energy and $k$-resolution, compared to that of the more bulk sensitive regime at higher photon energy. 
For either of these two regimes, $k$-space coverage and tunability are advantages compared to the laser ARPES regime that possesses both bulk sensitivity and the highest resolutions.
In this study, high contrast differences in valence band and Si 2$p$ core level spectroscopy have been used to spatially discriminate Si- and U-terminated surface domains of c-axis cleaved \URS.  
The identification of non-bulk surface states, 
essential for achieving the ultimate goal of determining the bulk electronic structure,
is made from this comparison of the two surface terminations assisted by response to surface modifications (aging, alkali deposition, and hydrogen dosing), and by comparison to bulk and surface slab theory calculations.  

\textbf{Surface States.}
The \SiT surface is found to host numerous $k$-dependent surface states arising from the entire top four-layer U-Si-Ru-Si half unit cell.
These surface states are the result of p-type energy shifts coming from the lack of a top U layer of charge and from top-site Si-dangling-bond orbitals.
The most distinct surface state is a strong intensity Si $p_z$-character 4-fold symmetry square hole-pocket at \Mbar\ (X-point zone boundary region) where a bulk-projected gap occurs in the non-$f$ bands. 
It is accompanied by a narrow $f$-state at \EF, primarily observed at the $f$-resonance condition, confined to the interior of the X-point hole surface state due to interaction between a fourth-layer U $f$-orbital and the top-layer Si $p_z$ orbital. 
At \Gbar, another very distinctive and intense hole-band surface state with second layer Ru d$_{x^2-y^2}$-character reaches a maximum energy of $\sim$30 meV below \EF, and its ubiquitous presence in the high-resolution ARPES literature identifies the \SiT surface being studied. 
Also identified from surface slab theory is the surface-region origin of another ubiquitous hole-pocket crossing \EF\ outside of the strong surface state below \EF.  

In contrast, the \UT surface, where the sole surface state tends to be a semi-localized broad $k$-independent U 5$f$ weight, provides the most bulk-like valence band and Si 2$p$ spectra. 
However the bulk valence band features, although notably stronger in U 5$f$ intensity than for the \SiT surface, tend to be sharper without this interfering surface U $f$-weight, and with the Si-Ru-Si overlayer protection of the next subsurface U layer.
Hence for the study of the bulk electronic structure in the low photon energy regimes, the \SiT surface is still preferred \textit{when} the bulk states are separable from the surface states. 
The complex X-point region for the \SiT surface, discussed in section \ref{xpt}, provides a counter-example for when such separation is not possible, and the quantitative determination of the Fermi-edge crossing of the shallow bulk electron band at the X-point is left ambiguous.

\textbf{f-renormalization.}
A key take away from this study is an appreciation for the significant $f$-band energy renormalization in \URS.
Comparison of the ARPES measurements to a DFT+DMFT bulk band calculation, with $\sim$5$\times$ overall energy renormalization relative to DFT, provided the impetus for explaining the quantitative discrepancy between DFT band theory FS and bulk sensitive SX-ARPES-derived FS in the PM phase.  In particular,  the measured Fermi-edge intensity patterns of streaks of $f$-weight along bulk \G-Z and X-Z directions, observed at high photon energy where U 5$f$ cross section is dominant, is shown to result from an inability to resolve the details of the heaviest effective mass $f$-bands that span large momentum ranges very close to \EF\ between large \kF\ and small \kF\ non-$f$ bands, as observed in $f^0$ \TRS.  
A  larger $\sim$10$\times$ energy renormalization scaling of the DFT energies at the X-point is shown in section \ref{xpt} to be consistent with both the band dispersion curvature of a shallow electron band crossing \EF, and the presence of another bulk band minimum 20 meV above \EF.
An even larger $\sim$20$\times$ narrower energy scaling of the zone-folded DFT band structure at \Gbar\ is required to make best agreement to the high-resolution ARPES $M$-band feature in section \ref{zonefold}.  
Such large $f$-band energy renormalizations then also bring the DFT $\sim$50 meV binding energy extended  flat-band region of the bulk N-point into sharp focus as being thermally active in the 20K temperature range of the HO transition.

\textbf{Bulk Fermi Surface.}
After identification of \SiT surface states and consideration of the larger energy renormalization of the $f$-bands relative to DFT predictions, the DFT bulk Fermi surface predictions in \textit{high symmetry} planes are largely confirmed by ARPES from various VB maps in the moderate to high-resolution energy regimes.  
Following the band index labeling established in figure \ref{trs}, we observe that the two small Z-point hole pockets (bands 2,3) in \TRS\ are pushed below \EF\ with the introduction of the U 5$f$ states as observed in figure \ref{vb150}.
In the same figures, the two distinct light mass hole-like bands in \TRS\ with large momentum \kF\ relative to Z (bands 4,5), are then observed, with sufficient energy resolution along Z-X-Z, to be transformed to have much smaller \kF\ values relative to Z via very heavy $f$-band-mass curvature consistent with the DFT+DMFT calculation.    
A more precise confirmation of the theory band 4 hole-band \kF\ comes from the outer edge of $M$-band measured by the high-resolution ARPES studies compared to the zone-folded DFT+DMFT bands in figure \ref{zfold}.  
Similarly, the inner part of the experimental $M$-band feature correlates well to the theory size of band 7 small electron pocket at the \G\ point.
High-resolution ARPES mapping of the zone-folded HO phase FS at 1K at a high symmetry 50 eV photon energy, observing 'petals' along (100) and a gapped FS along (110) \cite{Bareille2014}, provides a consistent picture with the DFT predictions of the size and shape of the band 5 (Z-point large hole) and band 6 (\G-point large electron) FS contours in the non-zone-folded PM phase.  
Lack of a precise quantitative determination of the FS size and depth of the bulk X-point shallow electron band (also band index 6) has been discussed in section \ref{xpt}.

Quantification of bulk FS contours shapes in the c-axis direction via photon-dependent \kxkz mapping is largely limited by the surface sensitivity and resultant $k_z$-broadening of features.   
Nevertheless, fuzzy intensity features in the polarization-dependent normal emission FS maps in figure \ref{gss} suggest a qualitative match to the DFT predictions for bands 4, 6 and 7, and with higher intensity at the crossover from hole-like to electron-like FSs intermediate between vertical Z- and \G-points.
The one critical exception to all this DFT FS agreement comes from the \kxkz map in figure \ref{npt}, where 
narrow band quasiparticle states at \EF\ are observed to be located significantly outside the DFT-predicted corners of the the band 5 FS contour.  These QP states are located symmetrically about the bct BZ N-point along the Z-N-Z line, where DFT predicts the existence of shallow binding energy necks connecting between the band 5 FSs in adjacent BZs.
Within a single BZ parallel to (100), these states also exhibit a 0.6a* nesting vector separation, suggestive that they are the true non-high-symmetry locus of $f$-states responsible for the neutron scattering spin-excitation Q$_1$ incommensurate momentum vector.

\textbf{Temperature Dependence.}
The ARPES $T$-dependence of the N-point region in section \ref{npt}, as well as of the \Gbar\ 
region in section \ref{gzpt}, exhibit evidence of a non-abrupt transition regime above \THO\ and below 30K.  
In particular, the T-evolution of the N-point electronic structure, which begins even above the Kondo coherence temperature of $T$*$\sim$50-70K, is observed to develop into the sharp narrow band QP states at \EF\ just above \THO. 
It is also noted that the shallow binding energy states at the N-point inherently possess a Q$_0$ commensurate nesting property, defined by the PM-phase Brillouin zone, which  
may provide a thermally-activated competition to the other Q$_0$ FS nesting condition between the top and bottom edges of the DFT band 5 and 6 FS sheets. 
Fuller theoretical and experimental examination of the N-point electronic structure illustrating its great potential for explaining key aspects of the hidden order state transition in \URS, and its relation to other ordered ground states is to be presented elsewhere \cite{Denlinger2022}.

\section*{Acknowledgements}
Research used resources of the Advanced Light Source, which is a US Department of Energy (DOE), Office of Science User Facility under contract no. DE-AC02-05CH11231.
Research at U. Michigan was supported by the US DOE under contract no. DE-FG02-07ER46379.
Research at UC San Diego was supported by the US DOE, Office of Basic Energy Sciences, Division of Materials Sciences and Engineering, under Grant No. DEFG02-04-ER46105 (single crystal growth) and the US National Science Foundation (NSF) under Grant No. DMR-1810310 (physical properties measurements).  Research at Rutgers U. was supported by the US NSF under Grant No. DMR 1709229. Research at the Catholic University of Korea (CUK) was supported by the National Research Foundation of Korea under Contract No. 2019R1A2C1004929. KK was supported by the internal R\&D program at KAERI (No. 524460-21). Identification of commercial equipment does not imply recommendation or endorsement by NIST.

\section*{ORCID iDs}

J. D. Denlinger  https:/orcid.org/0000-0001-7645-1631\\ 
J.-S. Kang		https:/orcid.org/0000-0002-3012-7673\\  
L. Dudy		https:/orcid.org/0000-0002-4204-5405\\ 
J. W. Allen	 	https:/orcid.org/0000-0001-7378-019X\\ 
Kyoo Kim		https:/orcid.org/0000-0002-7305-8786\\ 
K. Haule	         https:/orcid.org/0000-0002-2284-1551\\ 
J. L. Sarrao	https:/orcid.org/0000-0002-9357-661X\\ 
N. P. Butch       https:/orcid.org/0000-0002-6083-8388\\  
M. B. Maple	https:/orcid.org/0000-0002-5909-6057\\  

\section*{References}


\end{document}